\documentclass[fleqn,usenatbib]{mnras}

\newcommand{\flux}{\,erg\,s$^{-1}$\,cm$^{-2}$} %
\newcommand{\lumi}{\,erg\,s$^{-1}$} %
\newcommand{\gs}{\,g\,s$^{-1}$} %

\usepackage[T1]{fontenc}
\usepackage{ae,aecompl}

\usepackage{graphicx}	
\usepackage{amsmath}	
\usepackage{amssymb}	

\usepackage[symbol]{footmisc}
\usepackage{fp}
\usepackage{multirow}
\usepackage{footnote}

\title[V1298 Tau exoplanet evaporation]{X-ray irradiation and evaporation of the four young planets around V1298 Tau}

\author[K. Poppenhaeger et al.]{
K. Poppenhaeger,$^{1, 2}$\thanks{E-mail: kpoppenhaeger@aip.de}
L. Ketzer,$^{1, 2}$
and M. Mallonn$^{1}$
\\
$^{1}$Leibniz Institute for Astrophysics Potsdam (AIP), An der Sternwarte 16, 14482 Potsdam, Germany\\
$^{2}$Universit\"at Potsdam,  Institut f\"ur Physik und Astronomie,  Karl-Liebknecht-Stra\ss e 24/25, 14476 Potsdam, Germany}

\date{Accepted 2020 May 18. Received 2020 May 12; in original form 2020 March 2.}

\pubyear{2015}

\begin{document}
\label{firstpage}
\pagerange{\pageref{firstpage}--\pageref{lastpage}}
\maketitle

\begin{abstract}
Planets around young stars are thought to undergo atmospheric evaporation due to the high magnetic activity of the host stars. Here we report on X-ray observations of V1298 Tau, a young star with four transiting exoplanets. We use X-ray observations of the host star with Chandra and ROSAT to measure the current high-energy irradiation level of the planets, and employ a model for the stellar activity evolution together with exoplanetary mass loss to estimate the possible evolution of the planets. We find that V1298 Tau is X-ray bright with $\log L_X$\,[\lumi ] $=30.1$ and has a mean coronal temperature of $\approx 9$ MK. This places the star amongst the more X-ray luminous ones at this stellar age. We estimate the radiation-driven mass loss of the exoplanets, and find that it depends sensitively on the possible evolutionary spin-down tracks of the star as well as on the current planetary densities. Assuming the planets are of low density due to their youth, we find that the innermost two planets can lose significant parts of their gaseous envelopes, and could be evaporated down to their rocky cores depending on the stellar spin evolution. However, if the planets are heavier and follow the mass-radius relation of older planets, then even in the highest XUV irradiation scenario none of the planets is expected to cross the radius gap into the rocky regime until the system reaches an age of 5 Gyr.
\end{abstract}




\section{Introduction}

Exoplanets are very common around cool stars, with roughly one expected exoplanet per star \citep{Dressing2013}. How exoplanets evolve over time is a key question to understand the range of exoplanet properties we observe today, and to determine how stable exoplanetary atmospheres can be over long time scales.

Making use of the ever growing numbers of known transiting exoplanets, \citet{Fulton2017} detected a statistically significant gap in the regime of small exoplanets, manifesting as a valley or gap in the radius distribution at about 1.8 Earth radii. Using asteroseismic stellar parameters to refine the planetary radii, \citet{VanEylen2018b} showed that this gap has a dependence on the orbital period of the planet; this supports an interpretation that many of the small exoplanets are indeed evaporated cores of former larger planets with gaseous envelopes. While some migration in orbital distance can also be expected for exoplanets especially when the protoplanetary disk has not fully dissolved yet, the presence and slope of the gap suggests that evaporation may be main driver for its existence. However, other mass-loss scenarios like core-powered mass loss could in theory also lead to the observed gap in the radius distribution \citep{2020Loyd}. 

The main driver for atmospheric mass loss of exoplanets is thought to be the X-ray and extreme UV (together, XUV) irradiation the planet receives from its host star \citep{1981Watson, 2003Lammer, 2004Lecavelier}. This stellar emission is driven by the magnetic dynamo, which transforms the stellar rotation into the stellar magnetic field (see for example the review by \citealt{2017Brun}). The magnetism manifests itself as a variety of directly observable phenomena called activity, such as coronal and chromospheric emission, flares, and starspots. It is well-studied that the magnetic activity decreases over time as the star spins down through the process of magnetic braking mediated by the stellar wind (see review by \citealt{Guedel2007})). By the time the star reaches solar age, its XUV emission has typically decreased by about three orders of magnitude.

This means that by the time we observe the majority of small exoplanets -- typically around old main-sequence stars due to the better planet detectability when stars are inactive -- the atmospheric evaporation which forms the radius gap is mostly finished. However, in recent years the number of detected exoplanets in close orbits around young stars has grown rapidly. 
Such discoveries of small transiting exoplanets in young stellar clusters make it possible to study the XUV environment of exoplanets which are still at relevant ages for significant atmospheric mass loss. Mainly fueled by the K2 and TESS missions, a growing number of young exoplanets have been discovered by now \citep{David2016, Mann2016, Obermeier2016, Gaidos2017, Mann2017, Pepper2017, Livingston2018, Rizzuto2018, Newton2019}. The fact that their host stars have been identified as members of young stellar clusters gives an age tag to the star-exoplanet systems. 

A particularly intriguing system is the four-planet system around the star V1298 Tau. This star is a member of Group 29, an association in the foreground of the Taurus-Auriga association with a likely age of 23\,Myr \citep{David2019a}. 
It hosts four transiting planets \citep{David2019b} in orbits between 8 and 60 days with radii between 5.6 and 10.3~R$_\oplus$ (see table~\ref{table:sys_parameters} for the system properties).
The youth of this systems means that the stellar X-ray activity is high and can be measured precisely with present-day X-ray telescopes.

In this work we report on spectrally resolved X-ray observations of the host star V1298 Tau and extrapolate the extreme UV and X-ray (XUV) irradiation received by the exoplanets in the system. We estimate the current mass loss rates as well as the expected mass loss evolution of the planets over time scales of Gigayears, using an energy-limited evaporation model that takes into account the possible stellar activity evolution tracks.

\begin{table}

\caption{Properties of the V1298 Tau system as provided by \citet{David2019a, David2019b}}

\label{table:sys_parameters}
\begin{tabular}{ll}
\hline \hline
Parameter & Value \\ \hline
\textit{Star:} & \\
Spectral type &  K0-K1.5  \\
Stellar age [Myr] & 23 $\pm$ 4 \\
$M_\star$ [$M_\odot$] & 1.101$^{+0.049}_{-0.051}$  \\
$R_\star$ [$R_\odot$] & 1.345$^{+0.056}_{-0.051}$  \\
$P_\mathrm{rot}$ [d] & 2.870 $\pm$ 0.022   \\
Distance [pc] & 108.5 $\pm$ 0.7  \\ [0.1cm] 
\textit{Planet c:} & \\
$P$ [d] & 8.24958 $\pm$ 0.00072 \\
$R_P$ [$R_\oplus$] &  5.59$^{+0.36}_{-0.32}$ \\
$a$ [AU] & 0.0825 $\pm$ 0.0013 \\ [0.1cm] 
\textit{Planet d:} & \\
$P$ [d] & 12.4032 $\pm$ 0.0015 \\
$R_P$ [$R_\oplus$] &  6.41$^{+0.45}_{-0.40}$ \\
$a$ [AU] & 0.1083 $\pm$ 0.0017  \\ [0.1cm] 
\textit{Planet b:} & \\
$P$ [d] & 24.1396 $\pm$ 0.0018 \\
$R_P$ [$R_\oplus$] &  10.27$^{+0.58}_{-0.53}$ \\
$a$ [AU] & 0.1688 $\pm$ 0.0026 \\ [0.1cm] 
\textit{Planet e:} & \\
$P$ [days] & 60$^{+60}_{-18}$  \\
$R_P$ [$R_\oplus$] & 8.74$^{+0.84}_{-0.72}$ \\
$a$ [AU] & 0.308$^{+0.182}_{-0.066}$ \\
\hline

\end{tabular}

\end{table}

\section{Observations and data analysis}

The system was observed in X-rays by the ROSAT and Chandra space telescopes.

\begin{figure*}
\includegraphics[height=0.3\textwidth]{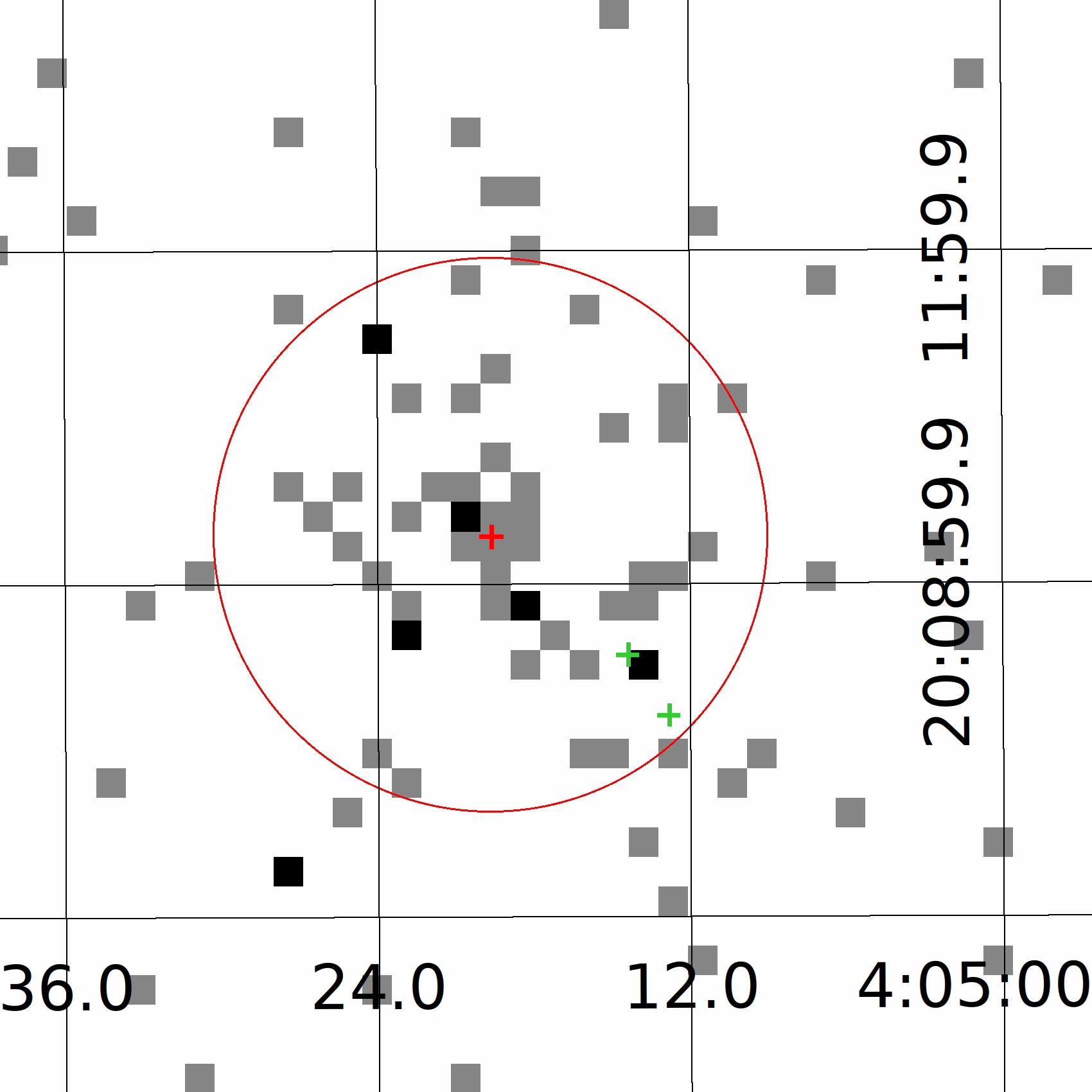}
\includegraphics[height=0.3\textwidth]{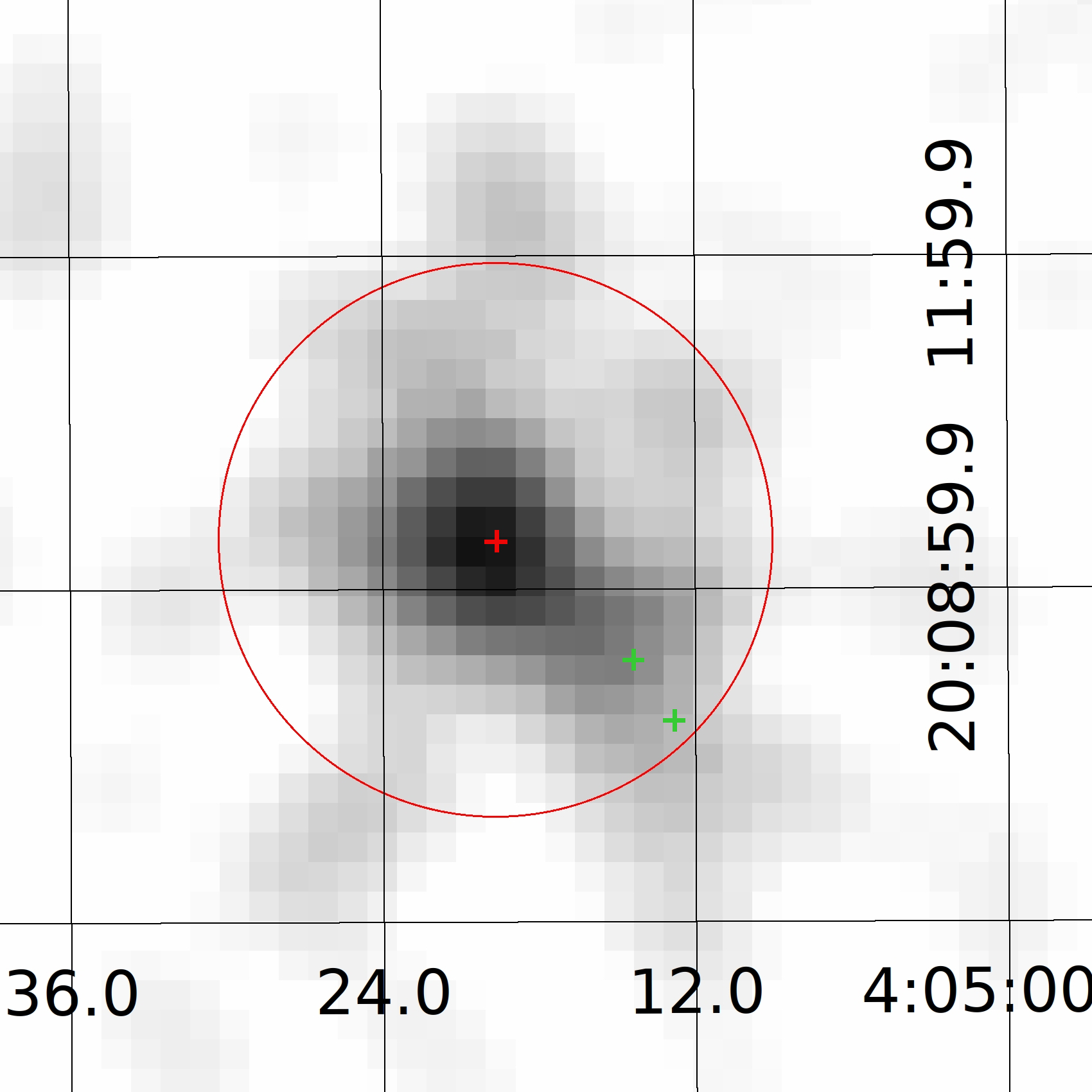}
\includegraphics[height=0.31\textwidth]{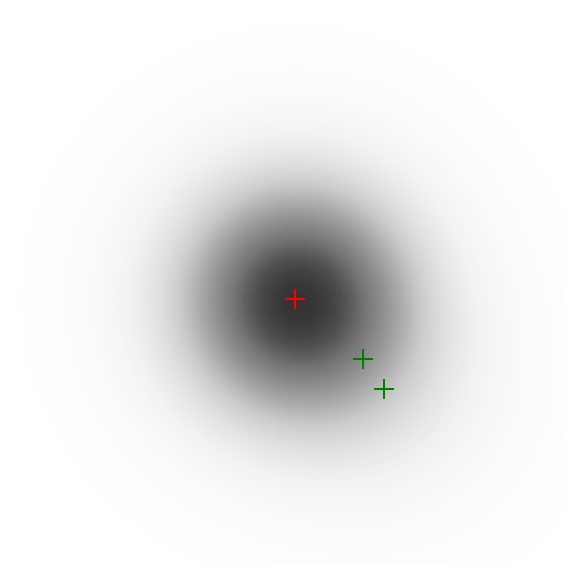}
\caption{ROSAT PSPC X-ray image of V1298~Tau, taken in 1991. The position of V1298~Tau is marked with a red cross, the positions of the two nearby young stars are marked as green crosses. The PSF extraction region with a radius of $150^{\prime\prime}$ is shown as a red circle. \textit{Left:} X-ray image with linear brightness scaling, binned spatially to a bin size of 16$^{\prime\prime}$. \textit{Middle:} Same, but smoothed by a Gaussian with a size of 4 bins. \textit{Right:} Fitted 2-D Gaussians to the positions of the three stars; the emission is dominated by X-rays stemming from the position of V1298~Tau.}
\label{fig:xrayimage_r}
\end{figure*}

\begin{figure}
\includegraphics[height=0.48\textwidth]{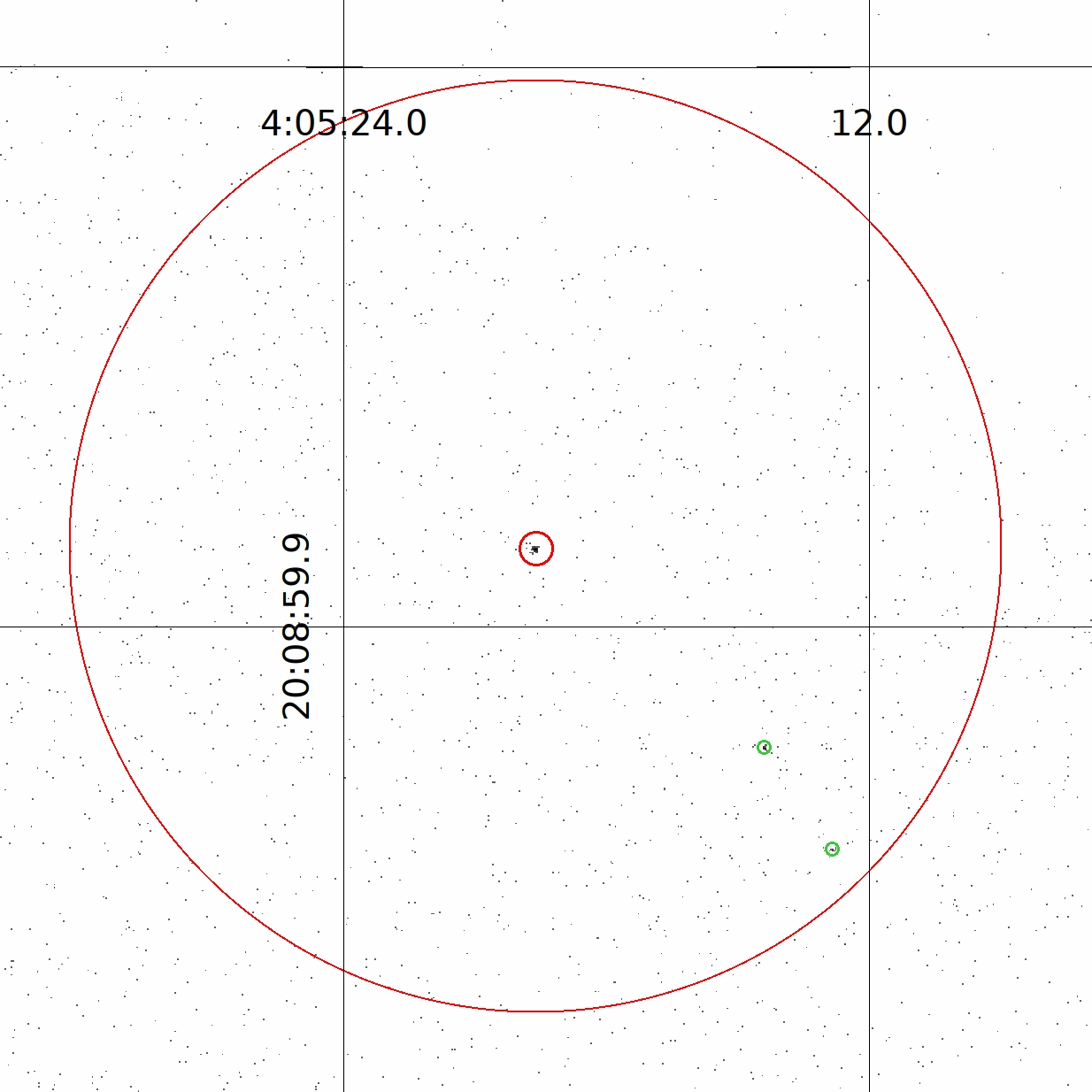}
\caption{Chandra ACIS-S X-ray image of V1298~Tau, taken in 2019. The position of V1298~Tau is marked with a small red circle, the positions of the two nearby young stars are marked with green circles, and the size of the ROSAT PSF is shown as a large red circle.}
\label{fig:xrayimage_ch}
\end{figure}

\subsection{ROSAT data}

The star V1298~Tau was observed with the ROSAT satellite in its ROSAT All-Sky Survey (RASS). ROSAT is an X-ray space telescope that was in operation from 1990 to 1994 \citep{Truemper1982, Aschenbach1988}, and performed the RASS for seven months during 1990 and 1991 in an energy band of 0.1-2.4 keV. The RASS consists of several scans of each part of the sky, with the individual visits of any given position being typically short, on the order of about 15--30 seconds, and accumulated exposure times of a few hundred seconds. V1298~Tau was observed in the RASS with PSPC-C detector.

We downloaded the archival X-ray data from the ROSAT archive\footnote[1]{\url{https://heasarc.gsfc.nasa.gov/FTP/rosat/data/pspc/processed\_data/}}. We used the \textit{xselect} software, which is part of NASA's HEASARC software package, to analyze the data. Specifically, we used the associated exposure map to determine the accumulated exposure time collected at the position of V1298~Tau, which was 297 seconds. We then defined a circular source region at the nominal position of the star with an on-sky radius of $150^{\prime\prime}$. This is motivated by the size of ROSAT's point-spread function (PSF), which varies considerably in width over the field of view; for accumulated RASS observations, a circular region of radius $150^{\prime\prime}$ contains about 90\% of the source flux \citep{Boese2000}. We also selected a background region free of obvious X-ray sources with similar exposure time and in the vicinity of the star. We opted for a background radius of $800^{\prime\prime}$ to obtain a more accurate determination of the background count rate to be expected in the source region.

We proceded by extracting the source and background region photon event lists and CCD spectra again for the source and background region using NASA's \textit{xselect} data analysis software. The spectra were grouped  to bins of at least five counts to avoid empty energy bins; the Cash statistic \citep{Cash1979} was used for spectral fitting with the \textit{Xspec} software.

\subsection{Chandra data}

Chandra \citep{Weisskopf2000} carries as one of its instruments the Advanced CCD Imaging Spectrometer (ACIS). It provides high spatial resolution with a PSF FWHM of 0.42$^{\prime\prime}$ at boresight, i.e.\ much higher than the spatial resolution of ROSAT. Its nominal energy range is 0.245 to 10 keV.

We obtained Chandra data for V1298 Tau using the ACIS-S detector in non-grating mode with an exposure time of 1.04\,ks on Nov.\ 17, 2019 (ObsID 22913). The image of the target was placed on one of the back-illuminated chips of ACIS-S, which provide slightly better energy resolution (approximately 70 eV FWHM for photon energies up to 2 keV) than the front-illuminated chips. ACIS-S provides a nominal energy sensitivity between 0.245 and 10 keV, but the sensitivity to X-ray photons below 0.8 keV has become very low over the life of the telescope. 

We used the CIAO software version 4.11 to reduce the Chandra data, employing the standard data analysis steps outlined in the CIAO user guide. We extracted light curves and CCD spectra for V1298 Tau, as well as for two other stars in V1298 Tau's vicinity. We chose a source extraction radius of 2$^{\prime\prime}$ radius and a background region of 60$^{\prime\prime}$ radius. Again, spectra were grouped to bins of at least five counts.

\subsection{Updated planetary ephemerides}

\begin{figure}
\includegraphics[width=0.48\textwidth]{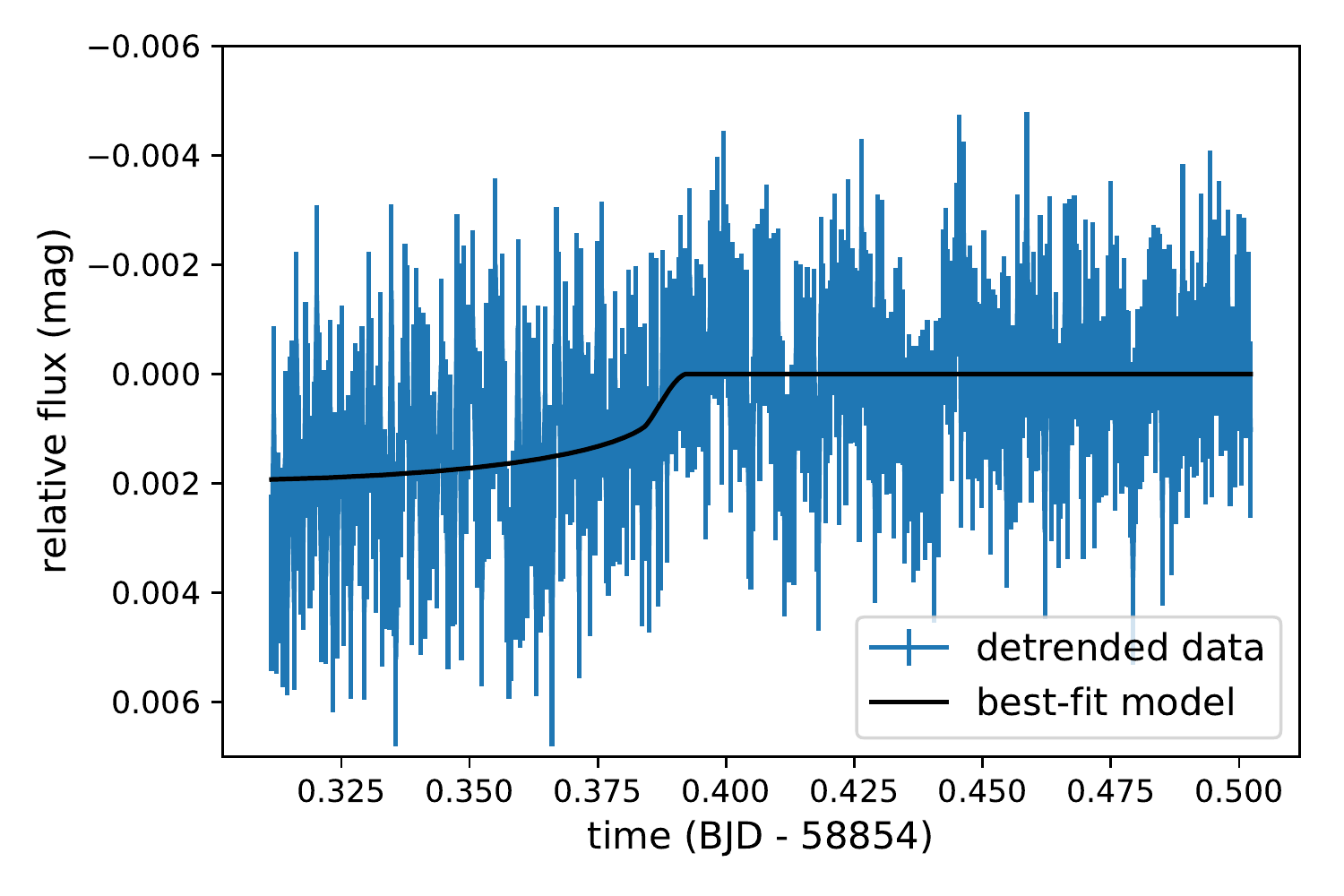}
\caption{Optical light curve of V1298 Tau during the transit of its innermost planet c. The data show the STELLA/WiFSIP observation after removal of the detrending function. The solid black line shows the best-fit transit model, indicating a transit time earlier than predicted by about 3.5~hours.}
\label{fig:optical_lc}
\end{figure}

While the transit depths of the four planets are small and are not expected to significantly alter the X-ray flux of the star during transits, we still report here the timing of the observations with respect to the orbital phases of the planets. As the system was observed with Kepler-K2 in 2015 \citep{David2019a}, the uncertainties in the orbital parameters have grown considerably since then. Typical transit midpoint uncertainties have increased to between three and five hours for observations in late 2019 (i.e.\ the epoch of the Chandra observation) for the innermost three planets. The outermost planets has an uncertain period measurement, so that its current ephemeris is unknown.

We have performed ground-based transit observations with the STELLA telescope \citep{Strassmeier2004} and its wide-field imager WiFSIP \citep{Weber2012} during the night of January 5, 2020 in order to cover a transit of the innermost planet $c$. The data were reduced following standard procedures employing a customised pipeline described in detail in \cite{Mallonn2015}. In brief, we used the publicly available software SExtractor for aperture photometry, and selected the aperture size that minimied the scatter in the light curve. The same criterion was applied to chose the ensemble of comparison stars for differential photometry. The light curve was analyzed with the transit modeling tool JKTEBOP \citep{Southworth2004}, keeping all relevant transit parameters fixed to the values derived by \cite{David2019b}. The parameters free-to-fit were the transit mid time and the three coefficients of a second order polynomial over time for light curve detrending. We found that the transit occurred earlier than expected from the \cite{David2019b} ephemeris by about 3.5 hours. Therefore, our optical transit observation started only at about transit mid time and lacks the transit ingress (Fig.~\ref{fig:optical_lc}).

We calculate that the start time of the Chandra observation occurred about 5.5 hours after the updated transit egress of planet c. With respect to planets d and b, the Chandra observation also took place well outside of those transits, even taking into account the current midpoint uncertainty of up to 5 hours for these planets. The uncertainty of planet e's mid-transit time is so large that we cannot determine whether the Chandra observation overlapped with it or not. 

The transit midpoint uncertainties for the three inner planets are of the order of one day for the ROSAT observing epoch. The accumulated ROSAT data covers a time period of almost 2 days, which may contain short exposures that were collected during a transit of one of the planets. We do not expect a significant influence of  a potential planetary transit on the ROSAT or Chandra data, since the transit depths are small with less than 0.5\% in the optical.

\section{Results}

\subsection{X-ray detection of V1298 Tau}

The ROSAT All-Sky source catalog lists a detected X-ray source near the nominal position of V1298~Tau (corrected for the known proper motion of the star to match the ROSAT observing epoch), with a spatial offset of $16.1^{\prime\prime}$ from its nominal position. This offset is not unusual given ROSAT's broad PSF. The detected count rate is 0.16 cps \citep{ROSATbrightsourcecatalog}, and a cursory calculation using V1298~Tau's distance of $108.2\pm0.7$\,pc \citep{GaiaDR2, Bailer-Jones2018} and a counts-to-flux conversion factor of $9.423\times10^{-12}$\,erg/cm$^2$/count\footnote[2]{Conversion factor derived according to \citet{Schmitt1995} using the hardness ratio of HR$_1$ = 0.21 from the ROSAT catalogue.} yields an estimate of $L_X = 2\times10^{30}$\,\lumi. This is close to the highest levels of X-ray luminosities observed for cool stars in general, and typical for very young stars like V1298~Tau \citep{Preibisch2005}.

We queried the Gaia DR2 archive for nearby stars brighter than $G=15$\,mag within a search radius of $150^{\prime\prime}$ around V1298 Tau's position. Most of the returned targets are located at distances much farther away than our target, by a factor of three or more; since field stars are intrinsically X-ray fainter than young stars, it is unlikely that those targets are origins for the detected ROSAT X-ray emission.

There are two other stars that have a similar distance as V1298~Tau, located at separations of  $\approx 100^{\prime\prime}$ and $130^{\prime\prime}$ from our target. These stars are HD 284154 (Gaia DR2 51884824140205824) and Gaia DR2 51884824140206720, which we will abbreviate as GDR2-5188 in the remainder of this publication.
The former was previously identified as a candidate member of the same young moving group as V1298 Tau \citep{Oh2017}. The latter is optically fainter and was not included in that candidate list, but has very similar distance and proper motion, so that all three stars may be members of the same young moving group, and as such intrinsically X-ray bright. We show an X-ray image of V1298 Tau's position taken with ROSAT in Fig.~\ref{fig:xrayimage_r}, indicating the position of all three stars. 

We also show the Chandra X-ray image of the same position in Fig.~\ref{fig:xrayimage_ch}, where the identification of the three stars is trivial due to Chandra's high spatial resolution. We find that V1298 Tau is clearly detected as the X-ray brightest source out of the three young stars, with a detected count number of 70 vs.\ 20 and 13 for HD 284154 and GDR2-5188, respectively. As a cursory check, we assume a hot coronal plasma temperature of about 10\,MK as appropriate for young stars and estimate an X-ray luminosity of $1.1\times10^{30}$\,\lumi\ for V1298 Tau from the detected number of counts and the exposure time using HEASARC's WebPIMMS tool\footnote{\url{https://heasarc.gsfc.nasa.gov/cgi-bin/Tools/w3pimms/w3pimms.pl}}; we will refine this estimate in the next section through spectral fitting. We can also confirm that there are no other X-ray sources detected with Chandra within the ROSAT PSF around V1298 Tau's position.


We can compare the X-ray brightness ratios between the three stars in the Chandra and ROSAT observations. For Chandra, this follows directly from the individually detected source counts. For the ROSAT data, we extracted an X-ray image with a spatial binning factor of 32; to guide the eye, we also extracted an image smoothed by a factor of four bins (see Fig.~\ref{fig:xrayimage_r}). We approximated the ROSAT PSF as a 2-dimensional symmetric Gaussian with standard deviation of $90^{\prime\prime}$, appropriate for the RASS PSF width, and fitted three Gaussians with free amplitude and centroids fixed to the positions of the three stars to the (unsmoothed) ROSAT image. The fit yielded that $\approx 90$\% of the X-ray photons come from the PSF centered on V1298~Tau's position, 10\% come from GDR2-5188, and the X-ray flux from the third star is compatible with zero.

The ROSAT brightness ratio between the stars of 90\% / 10\% / 0\% in terms of the added X-ray brightness of all three stars is roughly comparable to the X-ray brightness ratios in the Chandra data, namely 68\% / 19\% / 13\%, when one takes into account that the covered energy bands of the two telescopes are overlapping, but not the same, and the stars may display some intrinsic variability.


\subsection{Temporal variability}

\begin{figure}
\includegraphics[width=0.5\textwidth]{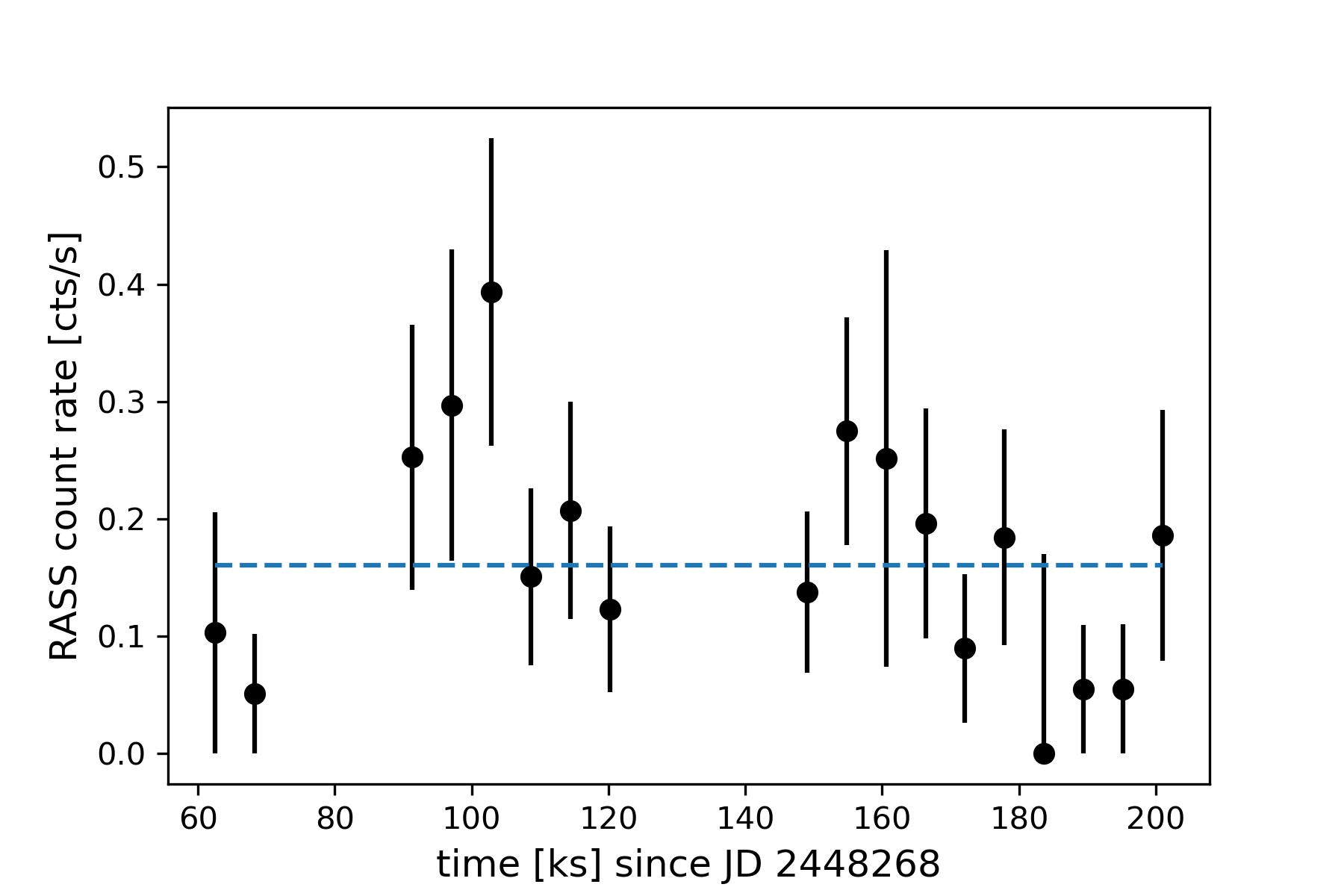}
\caption{ROSAT X-ray light curve of V1298~Tau in 1991, typical exposure time is around 20 seconds. The blue dashed line represents the mean count rate. }
\label{fig:rosatlc}
\end{figure}

\begin{figure}
\includegraphics[width=0.48\textwidth]{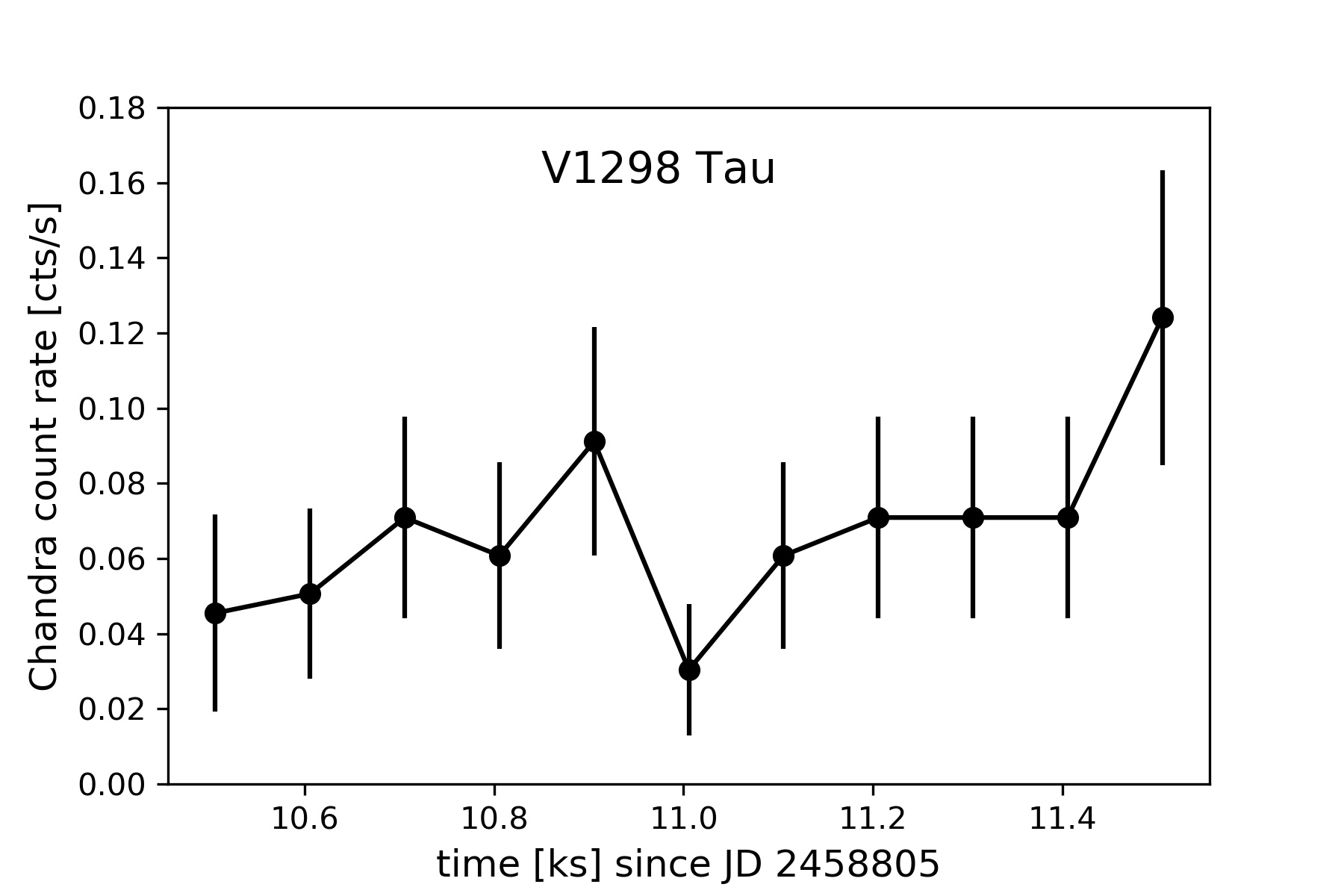}
\includegraphics[width=0.48\textwidth]{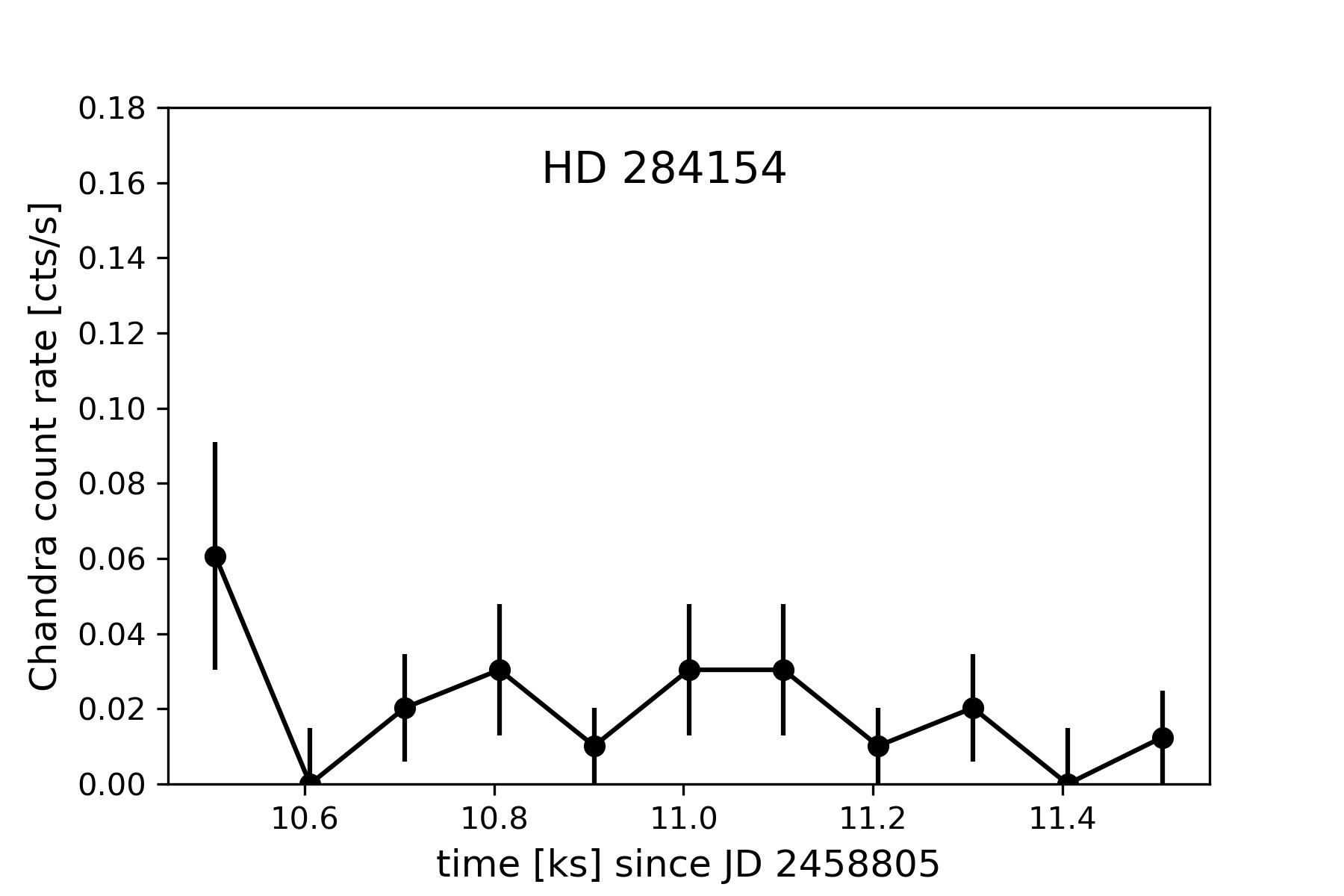}
\includegraphics[width=0.48\textwidth]{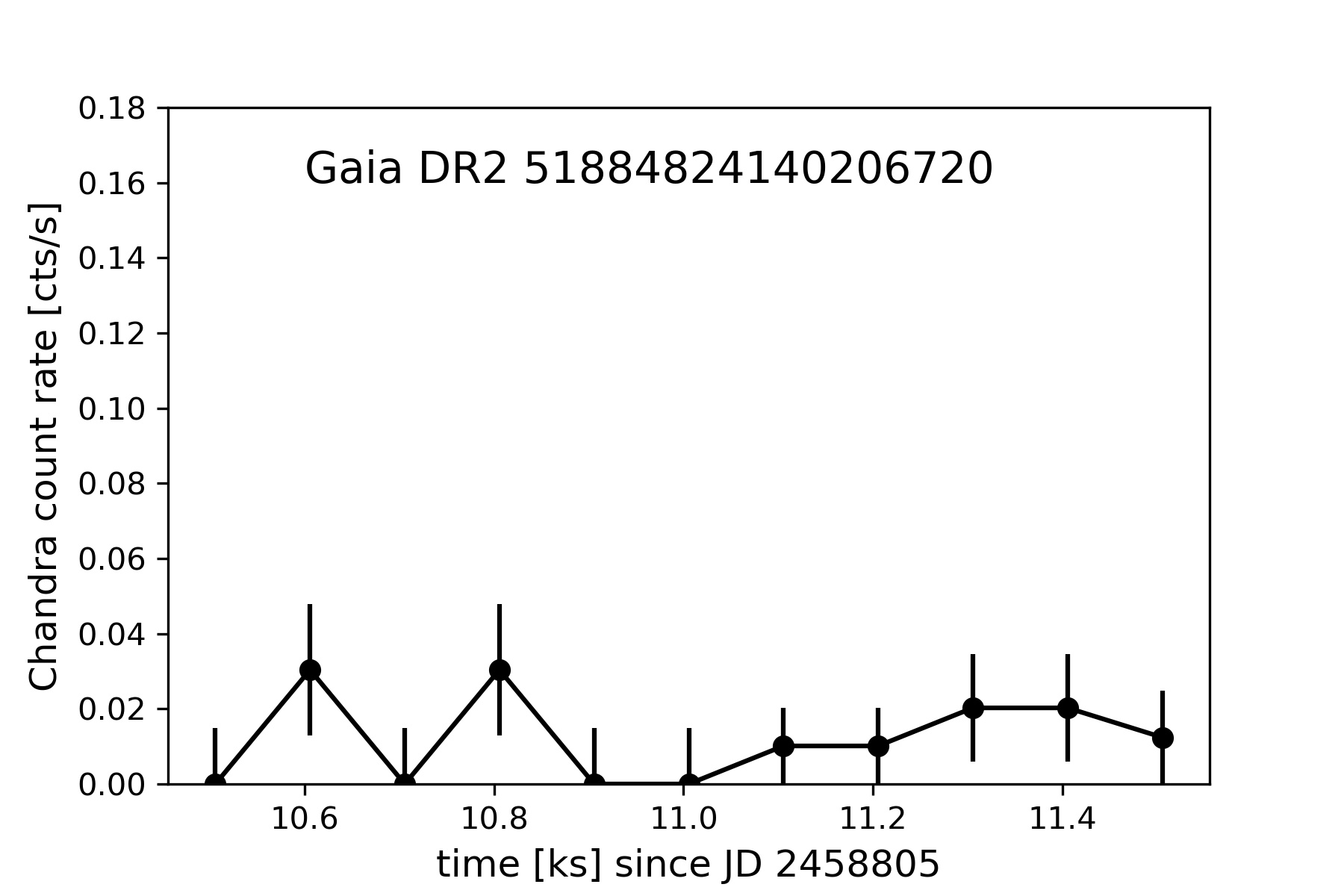}
\caption{Chandra X-ray light curves of V1298~Tau and the two nearby young stars in 2019 with a time binning of 100 seconds.}
\label{fig:chandralc}
\end{figure}

ROSAT scanned the position of V1298 Tau repeatedly over about two days, with individual exposure times of a few tens of seconds. As Fig.~\ref{fig:rosatlc} shows, the X-ray count rate in these individual exposures varies only mildly. No large excursions from the mean count rate, for example from flares, are observed. 

Within the Chandra observation, again no large flares are observed for V1298 Tau or any of the two other young stars, see Fig.~\ref{fig:chandralc}. The variability between time bins is consistent with uncertainties due to counting statistics. 

The absence of large flares in either of the observations means that the overall X-ray brightness ratios of the stars should indeed be similar, except for differences due to the covered energy range of the telescopes.

\subsection{Spectral fit and X-ray luminosity}

\subsubsection{V1298 Tau}

\begin{figure}
\includegraphics[width=0.45\textwidth]{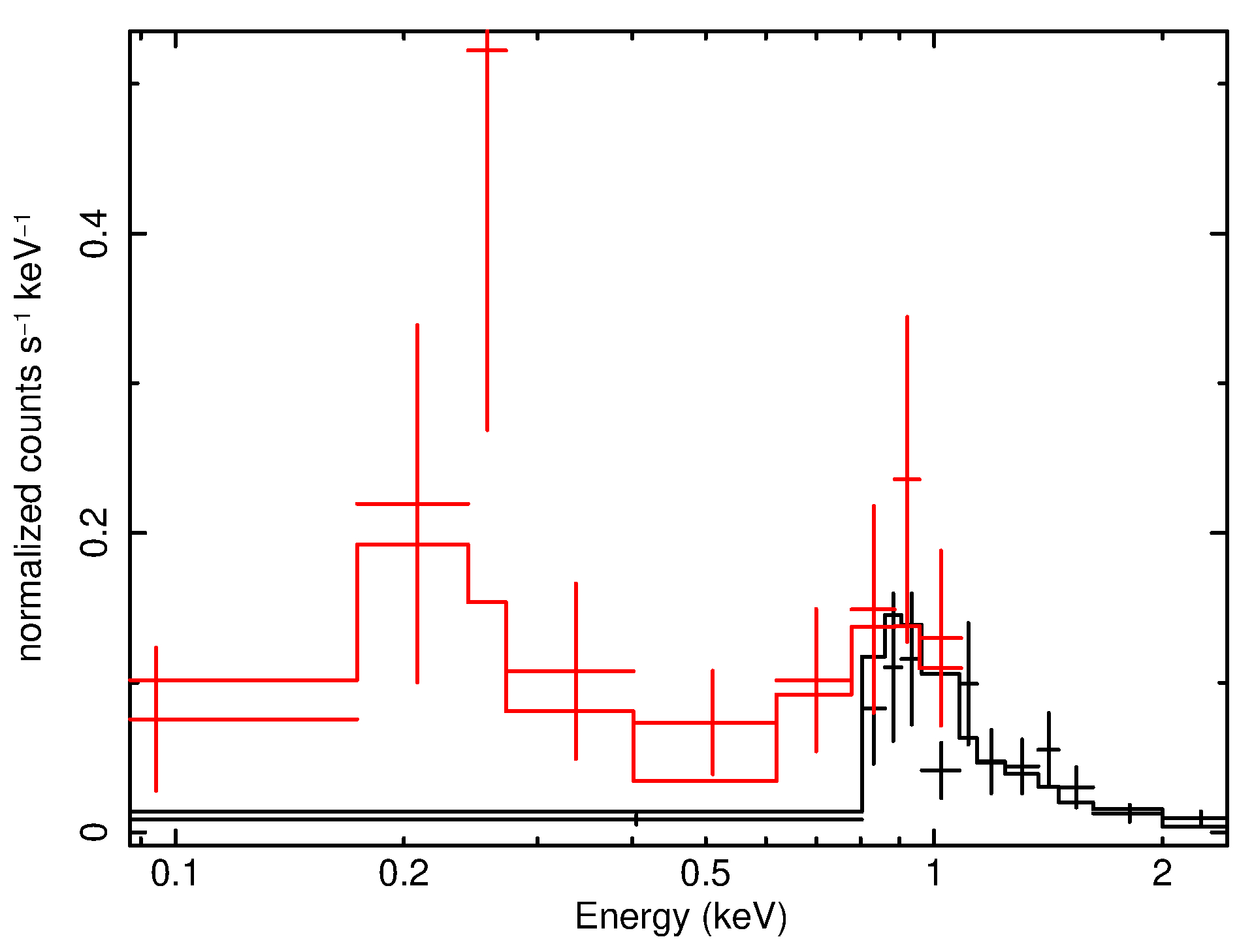}
\caption{ROSAT (red) and Chandra (black) X-ray spectra of V1298~Tau. The data points are shown as crosses with $1\sigma$ vertical error bars per energy bin. The best spectral fit is shown as a red and black line for the data sets from the two telescopes, respectively. }
\label{fig:xrayspec}
\end{figure}

To determine an accurate value for V1298~Tau's X-ray luminosity, we extracted CCD spectra from ROSAT and Chandra at V1298 Tau's position. The two data sets cover overlapping, but different energy ranges. ROSAT's low-energy boundary is 0.1 keV. While Chandra's low-energy sensitivity nominally extends down to 0.245 keV, in practice the effective area below 0.8 keV has become very low over the lifetime of the telescope. This means that any emission measure at low coronal temperatures and therefore softer photon energies than 0.8 keV are effectively not  probed by the Chandra data. Conversely, ROSAT's high-energy sensitivity ends at 2.4 keV, while Chandra's extends to 10 keV. For our source it turns out that there are almost no photons detected above 2.4 keV even in the Chandra data.

We show both the ROSAT and Chandra X-ray spectrum of V1298 Tau in Fig.~\ref{fig:xrayspec}. The different sensitivities of the two instruments become apparent immediately. 

We fitted the X-ray spectra with HEASRAC's Xspec software, using the APEC model appropriate for a coronal plasma \citep{Foster2012ATOMDB} and solar-like abundances from \citet{Grevesse1998}. The extinction towards V1298 Tau is low with $E(B-V)=0.024$, translating to a low gas column density $N_H$ of $\approx 1.6\times 10^{20}$\,cm$^{-2}$, which means that X-ray absorption by the interstellar medium is negligible for our target. We grouped the spectra to a minimum of five counts per bin and used the Cash statistic in Xspec's fitting process. 

In fitting the two spectra individually, we found that the fitted flux in the energy range where both instruments have decent effective area (0.8-2.4 keV) is similar for both observations with $5.3\times10^{-13}$\,\flux\ for ROSAT and $4.5\times10^{-13}$\,\flux\ for Chandra. We therefore decided to fit both spectra simultaneously with the same model to gain better energy coverage for V1298 Tau.

The result of the combined spectral fit is listed in Table~\ref{tab:fit}. The fitted coronal temperature is moderately high with a value of $kT=0.78$\,keV, corresponding to about 9 MK. The X-ray flux derived from the spectral fit is $9.2\times 10^{-13}$\,\flux\ for the native ROSAT energy band of 0.1-2.4 keV. The uncertainty in this flux was estimated with Xspec to be $\pm 1.0\times 10^{-13}$\,\flux\, i.e.\ about 11\% of the flux (given is the 68\% confidence interval). 
We extrapolated this to two other commonly used energy bands of 0.2-2 keV and 0.2-10 keV, yielding fluxes of $8.0\times 10^{-13}$\,\flux\ and $8.4\times 10^{-13}$\,\flux, respectively.
This places V1298~Tau's X-ray luminosity at $1.3\times 10^{30}$\,\lumi\ for the 0.1-2.4 keV band (equalling $1.1\times 10^{30}$\,\lumi\ and $1.2\times 10^{30}$\,\lumi\ for the 0.2-2.0\,keV and 0.2-10\,keV bands, respectively). 



Using the scaling laws between X-ray and extreme UV (EUV) emission of stars derived by \citet{Sanz-Forcada2010} and the ROSAT energy band as the input X-ray band, we estimate the combined XUV (X-ray and EUV) luminosity of V1298~Tau to be $L_{\mathrm{XUV}} = 6.3\times 10^{30}$\,\lumi\ for an EUV energy band of 0.01-2.4 keV.

\subsubsection{The two young stars HD 284154 and GDR2-5188}

The two nearby young stars do not have sufficient source counts in the Chandra observation to perform an adequate spectral fit for them. However, we can assess their hardness ratios and compare them to the hardness ratio of V1298 Tau. Specifically, we calculate the hardness ratio $HR = \frac{H-S}{H+S}$ with a soft band $S$ of [0.5--1.5) keV and a hard band $H$ of [1.5--5.0] keV. 

V1298 Tau, HD 284154 and GDR2-5188 then display hardness ratios of $-0.5\pm 0.1$, $-0.8_{-0.2}^{+0.3}$, and $-0.2\pm 0.3$, respectively. These are consistent with each other within their uncertainties, indicating roughly similar spectral shapes. We therefore estimate the X-ray luminosities of the two fainter stars by scaling V1298 Tau's X-ray luminosity down by a factor given by the X-ray count ratios between V1298 Tau and the respective other star. 

In this manner, we estimate HD 284154's and GDR2-5188's X-ray luminosity to be 0.29 and 0.19 times the X-ray luminosity of V1298 Tau, respectively. In absolute numbers, we estimate the X-ray luminosities of HD 284154 and GDR2-5188 in the 0.1-2.4 keV energy band to be $3.8\times10^{29}$\,\lumi\ and $2.5\times10^{29}$\,\lumi , respectively.


\begin{table}
\caption{Spectral fitting results for V1298 Tau from the combined Chandra and ROSAT data.}
\label{tab:fit}
\begin{tabular}{l l}
\hline\hline
Parameter & Value \\ \hline \\ [-0.3cm]
kT [keV] & $0.78\pm 0.07$ \\
EM [cm$^{-3}$]\footnote{Emission measure; corresponding to a fitted "norm" parameter in the APEC model of $(3.2 \pm 0.3)\times 10^{-4}$.} & $(4.5 \pm 0.4)\times 10^{52}$ \\ [0.2cm]
$F_{\mathrm{X,\,0.1-2.4\, keV}}$ [\flux] & $(9.2\pm 1.0)\times 10^{-13}$\\
$F_{\mathrm{X,\,0.2-2\, keV\,\,\,\,}}$ [\flux] & $(8.0\pm 0.9)\times 10^{-13}$\\
$F_{\mathrm{X,\,0.2-10\, keV\,}}$ [\flux] & $(8.4\pm 0.9)\times 10^{-13}$\\ [0.2cm]
$L_{\mathrm{X,\,0.1-2.4\, keV}}$ [\lumi] & $(1.3\pm 0.1)\times 10^{30}$\\
$L_{\mathrm{X,\,0.2-2\, keV\,\,\,\,}}$ [\lumi] & $(1.1\pm 0.1)\times 10^{30}$\\
$L_{\mathrm{X,\,0.2-10\, keV\,}}$ [\lumi] &$(1.2\pm 0.1)\times 10^{30}$ \\ [0.2cm]
$L_{\mathrm{XUV,\,0.01-2.4\, keV\,}}$ [\lumi] &$(6.3\pm 0.1)\times 10^{30}$ \\ 

\hline
\end{tabular}
\end{table}

\section{Discussion}

\subsection{V1298~Tau's activity evolution in the context of young stars}

We can place the X-ray emission level of V1298 Tau, as well as for the other two young stars, into the context of cluster stars of a similar age. The cluster NGC 2547 has a similar age of about 30 Myr and has been studied in X-rays with both ROSAT and XMM-Newton data \citep{Jeffries2006}. A comparison of the fractional X-ray luminosity $L_X/L_{bol}$ versus photometric stellar colours is particularly instructive, because it displays the lower activity state of higher-mass stars compared to lower-mass ones at a fixed given age.

We calculated the fractional X-ray luminosity and photometric colours of the three stars studied here as follows: We queried the Gaia DR2 archive for the approximated bolometric luminosities of the stars, which are given as 0.9, 4.0, and 0.2 for V1298 Tau, HD 284154, and GDR2-5188, respectively. We transformed the stellar $B_P-R_P$ colours from Gaia into $V-I_c$ colours according to the Gaia Data Release 2
Documentation release 1.2\footnote{\url{https://gea.esac.esa.int/archive/documentation/GDR2/}}, which yielded $V-I_c$ = 1.0, 0.7, and 1.6 mag, respectively.

\begin{figure}\includegraphics[width=0.5\textwidth]{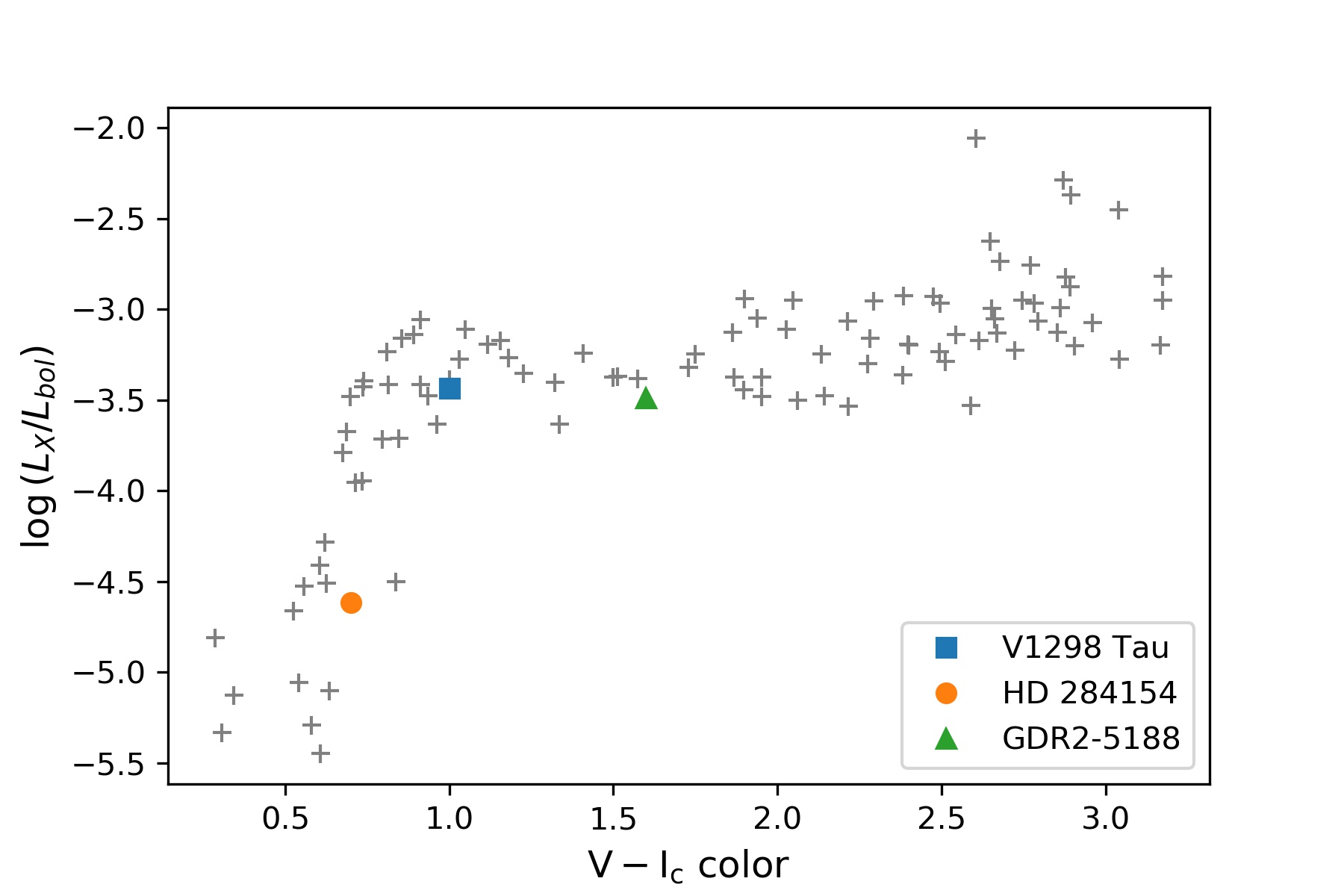}
\caption{Fractional X-ray luminosity ($L_X/L_{bol}$) of V1298~Tau and the other two young stars as a function of their $V-I_c$ colour, together with fractional X-ray luminosities of member stars of the 30 Myr old cluster NGC 2547 (cluster data from \citet{Jeffries2006}).}
\label{fig:cluster}
\end{figure}

The placement of the three stars studied here in context with the cluster NGC 2547 are shown in Fig.~\ref{fig:cluster}. All three stars fall within the range of typically observed fractional X-ray luminosities for stars of a similar age. This means that neither of the stars is particularly active of inactive for their age.

\subsection{Evaporation of the four planets}

We investigate the possible future evolution of the four young planets around V1298 Tau with respect to their atmospheric mass loss induced by the stellar X-ray and extreme UV irradiance. The goal of the following calculations is to show the breadth of possible planetary masses and radii at a mature system age. We will demonstrate that both the present-day masses of the four planets, which are currently still unknown, and the specific stellar activity evolution in the next few hundred million years have a strong influence on how the planets in the system evolve. We have made the python code for these calculations publicly available\footnote{\url{https://github.com/lketzer/platypos/}}.

We start by adopting the commonly used energy-limited hydrodynamic escape model \cite[see e.g.][]{OwenJackson2012, Lopez2012}:

\begin{equation}
\dot{M} = \epsilon \frac{(\pi R_{XUV}^2) F_{\mathrm{XUV}}}{K G M_{pl}/R_{pl} } = \epsilon \frac{3 \beta^2 F_{\mathrm{XUV}}}{4 G K \rho_{pl}}\,,
\label{eq1}
\end{equation}

where $F_{\mathrm{XUV}}$ is the flux impinging on the planet, $R_{pl}$ and $R_{XUV}$ are the planetary radii at optical and XUV wavelengths, respectively; we use $\beta = R_{XUV}/R_{pl}$ as a shorthand in the following. $M_{pl}$ is the mass and $\rho_{pl}$ the density of the planet, $\epsilon$ is the efficiency of the atmospheric escape with a value between 0 and 1, and $K$ is a factor representing the impact of Roche lobe overflow \citep{Erkaev2007}, which can take on values of 1 for no Roche lobe influence and $<1$ for planets filling significant fractions of their Roche lobes. For the present-day V1298~Tau system the $K$ factor takes on values of about 0.8 to 0.9 from the innermost to the outermost planet, with a slight dependence on the assumed masses of the planets which have not been measured yet.

\subsubsection{Planetary mass estimates}

Measuring masses of planets around highly active stars like V1298 Tau is challenging, as stellar activity can mask and distort the radial velocity signatures of planets. No masses for the V1298 Tau planets have been published so far. We therefore estimate their masses based on two different assumptions.

It is possible that the V1298 Tau planets follow the empirical mass-radius relationships observed for planets around older stars \citep{ChenKipping2017, Otegi2019_arXiv}. These relationships show two regimes, one for small rocky planets up to radii of about $2R_\oplus$ and one for larger planets with volatile-rich envelopes. The scatter is low in the rocky planet regime and larger in the gaseous planet regime: as core vs.\ envelope fractions may vary, there is a broader range of observed masses at a given planetary radius for those larger planets. It is noteworthy that the young planet K2-100b, which has an age of $\approx 700$ Myr based on the cluster membership of its host star \citep{Mann2017}, falls into the volatile envelope regime and follows the mass-radius relationship seen for older planets.

If this is correct also for the younger V1298 Tau planets, we can estimate their masses from the mass-radius relationship presented by \citet{Otegi2019_arXiv}; the relationship by \citep{ChenKipping2017} yields essentially the same results. The present-day planetary masses then span a range of approximately 26 to 69 $M_\oplus$. We refer to this assumption as the ''high-density scenario'' in the following. We note here that we consider this ''high-density scenario'' as somewhat less likely than the ''fluffy planet scenario'' described further below. This is due to considerations about the mutual Hill separations of the planets as reported by \citet{David2019b} in their section 3.1, where they show that if the V1298 Tau system follows the typically observed Hill separations in other multi-planet systems, then the combined mass of the two innermost planets should be $7^{+21}_{-5}\,M_\oplus$. The ''high-density scenario'' would yield a combined mass of approximately $59\,M_\oplus$ instead.

In contrast to the mass-radius relation of mature planets, very young planets may display enlarged radii as they are not dynamically settled yet. Simulations of planet formation and evolution (see for example \citealt{Mordasini2012}) show that in the age range of 10 to 50 Myr planets evolve from a wide range of possible radii at a given mass towards a more unified mass-radius relationship. If planets are formed under the so-called cold start scenario, i.e.\ with low initial entropy, their radii may not shrink very much as the planets age. However, if planets are formed according to the hot start scenario, i.e.\ with a high initial entropy, their radii may be considerably larger at young ages compared to older ages. The planet $\beta$ Pic b shows indications of having been formed under hot start conditions \citep{Snellen2018}. 

If this is also the case for the V1298 Tau planets, their current masses could be much lower than estimated by a mass-radius relationship valid for older planets. We approximate this scenario by using models of planets with a hydrogen/helium envelope on top of a 5 and 10 M$_\oplus$ core, using the tabulated models of \citet{LopezFortney2014}. They calculate radii for low-mass planets with hydrogen-helium envelopes on top of Earth-like rocky cores, taking into account the cooling and thermal contraction of the atmospheres of such planets over time. Their simulations extend to young planetary ages, at which planets are expected to still be warm and possibly inflated. \citet{LopezFortney2014} provide simple analytical fits to their simulation results, which we use to trace the thermal and photoevaporative evolution of the planetary radius over time. We refer to this as the ''fluffy planet scenario'' in the following.\\

\subsubsection{Evaporation parameters}

Different values for the efficiency parameter $\epsilon$ have been suggested in the literature, ranging from 0.4 \citep{Lalitha2018} down to 0.01 and even lower for high-mass planets like CoRoT-2b \citep{Salz2016b}. In contrast, the V1298 Tau planets are well below Jupiter size and can be expected to have moderately low masses (see above). 
\cite{Salz2016b} used hydrodynamic simulations of exoplanets in close orbits to estimate $\epsilon$. For planets of relatively low mass (and therefore low gravitational potential energy) and high irradiation level they reported efficiency values between 0.1 and 0.3. 
In our work, we choose an $\epsilon$ value of 0.1 for all of our mass-loss rate calculations; we refer the reader to section~\ref{limitations} for a discussion of limitations of the model.

The XUV radii of exoplanets have in some cases been found to be significantly larger than their optical radii from observations at UV and X-ray wavelengths \citep{Poppenhaeger2013}. We use here again an approximation by \citet{Salz2016b}, who derived a scaling law for the planetary XUV radius as a function of planetary gravitational potential and their XUV irradiation. Assuming planetary masses follow a mass-radius relationship also valid for older planets \citep{Otegi2020}, we find XUV radii for the four planets that are approximately 1.5 to 1.7 times larger than their respective optical radii. If the planets are fluffy, their gravitational potential is lower by half an order of magnitude than the sample simulated by  \citet{Salz2016b}; if we extrapolate their relationship for XUV radii to this regime, we find that the planetary XUV radii can be enlarged by factors of about 1.5 to 2.3 compared to the optical radius for the individual planets. This significant radius enhancement increases the calculated mass loss rates by a factor of about 2 to up to 5 compared to an XUV radius that is the same as the optical radius.

\subsubsection{Present-day mass loss rates of the planets}

The expected present-day mass loss rates of the planets depend sensitively on the assumed masses of the planets. We report the mass loss rates for the high density scenario and for the fluffy planet scenario, the latter one using two different potential core masses of 5 and 10 $M_\oplus$. We take into account the measured stellar X-ray luminosity and its extrapolation to the XUV wavelength band, as well as the Roche-lobe overflow factor $K$ \citep{Erkaev2007} and the planetary XUV radius as simulated by \citet{Salz2016b}. 

The XUV irradiation in the energy range of 0.01-2.4 keV at the planetary orbital distances is high compared to more mature exoplanet systems; we find $F_\mathrm{XUV,\,orbit} = 32.9, 19.1, 7.9, 2.4 \times 10^5$~\flux \, from the innermost to the outermost planet (i.e.\ planets $c$, $d$, $b$, and $e$).

The typical difference in expected mass loss rates differs by about an order of magnitude between assuming the high-density scenario and the fluffy planets scenario with a core mass of $5\,M_\oplus$ for the planets. The innermost planet $c$ yields an expected present-day mass loss of $2.5\times 10^{13}$\gs\ for the fluffy $5\,M_\oplus$ core mass scenario, $8.7\times 10^{12}$\gs\ for the fluffy $10\,M_\oplus$ core mass scenario, and $2.4\times 10^{12}$\gs\ for the high-density scenario. The derived expected mass loss rates for all planets and considered scenarios are listed in Table~\ref{tab:mass_evo}.





\subsubsection{Stellar activity evolution}

In order to investigate the atmospheric erosion that the four planets might undergo in the future, it is crucial to take into account the change in XUV flux received by the planets over time. Many studies of exoplanet evaporation approximate the stellar XUV evolution by using the average activity level of stars in a specific mass bin for well-studied clusters of different ages. This can be represented as a power law decrease in activity which sets in after some time scale during which the stellar activity stays constant at a saturation level \citep{Ribas2005, Jackson2012}. 

However, the rotational and therefore activity evolution of stars with similar mass in young clusters shows a strong spread, which manifests itself as the so-called slow and fast rotational sequences \citep{Barnes2003}. It is possible that a star spins down early and follows a low-activity track, or that it maintains its high rotation rate and activity for a longer time and spins down later. Specifically, the observed rotational evolution of stars in clusters has been interpreted as the stellar transition from fast to slow rotation happening quickly for individual stars, but at different stellar ages in the same cluster \citep{Garraffo2016}.  In the context of exoplanet irradiation, this was explored in simulations by \citet{Tu2015, Johnstone2015}. Their studies show that the saturation timescales may range from $\sim\,10$ to $\sim\,300\,$Myr for stars of the same mass. Whether a star follows a high- or low-activity track can make a significant difference for the evaporation of its exoplanets.

\begin{figure}\includegraphics[width=0.5\textwidth]{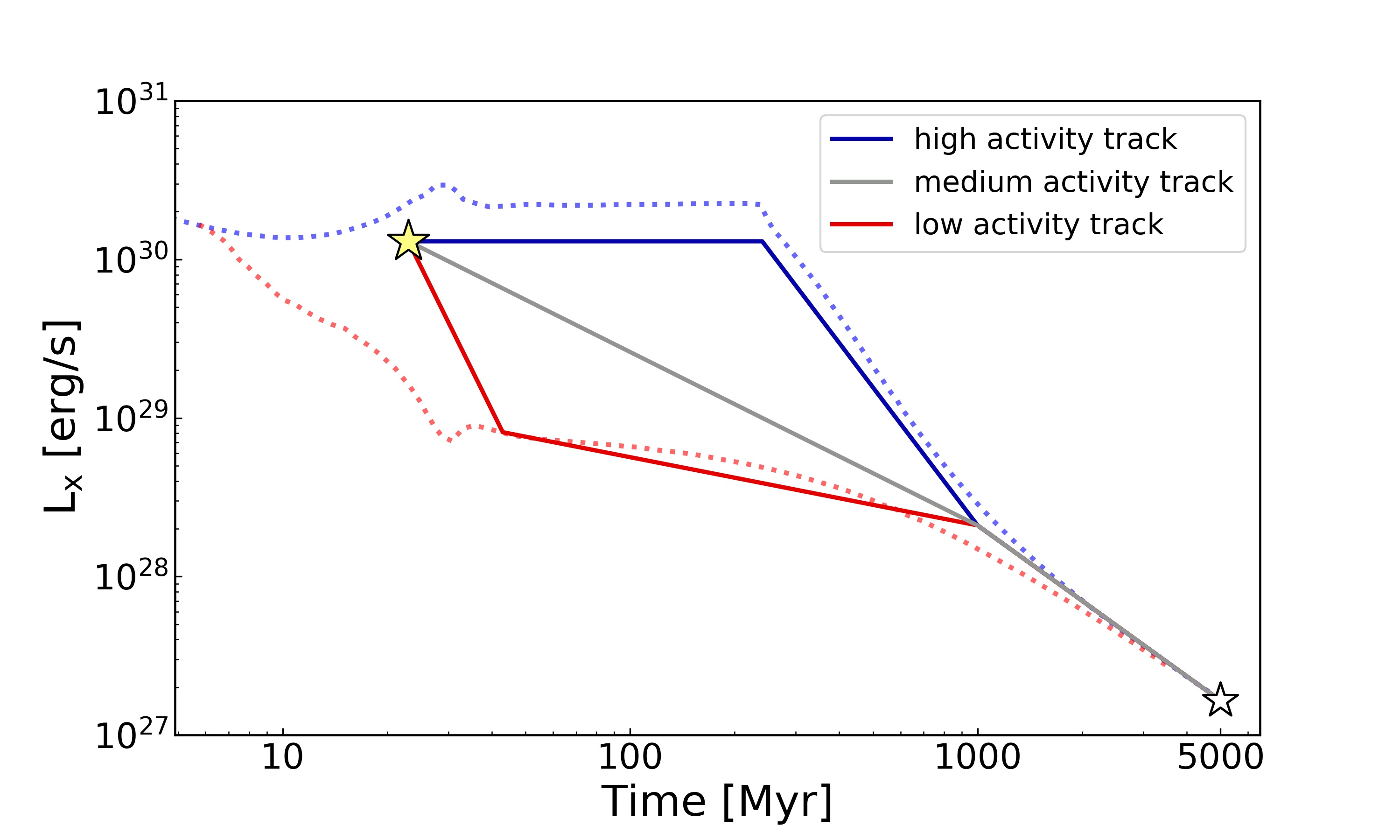}
\caption{High, low and medium activity tracks (blue, grey and red, respectively) showing the future $L_X$ evolution of V1298~Tau assumed in our calculations. The yellow star indicates the current X-ray luminosity of the system, while the white star marks the X-ray luminosity at the end of our calculation at 5~Gyr. The blue and red dotted model tracks (shown for comparison) are taken from \citep{Tu2015} and represent the X-ray evolution for an initially fast and slowly rotating solar-like star.}
\label{fig:evo_tracks}
\end{figure}

Inspired by \citet{Tu2015}, we use a simplified broken power-law model to approximate the solar-mass stellar activity evolution, which we show in Fig.~\ref{fig:evo_tracks}. Specifically, we define a high-activity track, where the star stays very active for a long time, and a low-activity track, where the spin-down and therefore the activity decrease happens early in the life of the star. For comparison with studies using an average activity evolution, we also define an average activity track. We let our high-activity track start at the current activity level of V1298 Tau; it is possible that other stars of the same age and mass may have an even higher activity level.

\subsubsection{Mass loss evolution of the planets}

We calculate the evaporation of the four planets for a variety of scenarios. We take into account the possible high-density and fluffy planet scenarios, as explained above, and further divide these cases into different stellar activity evolution scenarios.

In our calculations we compute the momentary mass loss rate of each planet according to equation \ref{eq1}, having calculated $R_{XUV}$ and $K$ for the fluffy and the high-density planet scenario, respectively. We adjust the gaseous envelope mass of the planet and calculate its new radius, again using either the fluffy or high-density planet assumption. 

To model the four young planets in the ''fluffy planet scenario'', we make use of the simulation results from \citet{LopezFortney2014} and assume core masses of either 5 or 10 M$_\oplus$ for all planets in the system. Coupled with the mass-loss formalism introduced above (see equation \ref{eq1}), we allow for the atmospheric photoevaporative mass loss of the envelope and take into account the thermal contraction of the planets in the model by \citet{LopezFortney2014}. For the fluffy planet scenario, we stop the simulation and assume the complete gaseous envelope has been evaporated once the planetary radius matches the core radius. 

In the ''high-density scenario'' we assume that the planetary radii change according to the mass-radius relation for the more evolved, volatile-rich planet population, considering the decrease in planetary masses as a result of the XUV induced mass loss. If the planetary size reaches $2R_\oplus$, which we assume to be the upper end of possible core sizes based on the location of the exoplanet radius gap, we stop the simulation and assume the complete gaseous envelope has been evaporated.

\begin{figure}
\includegraphics[trim=30 25 30 0,clip,width=0.50\textwidth]{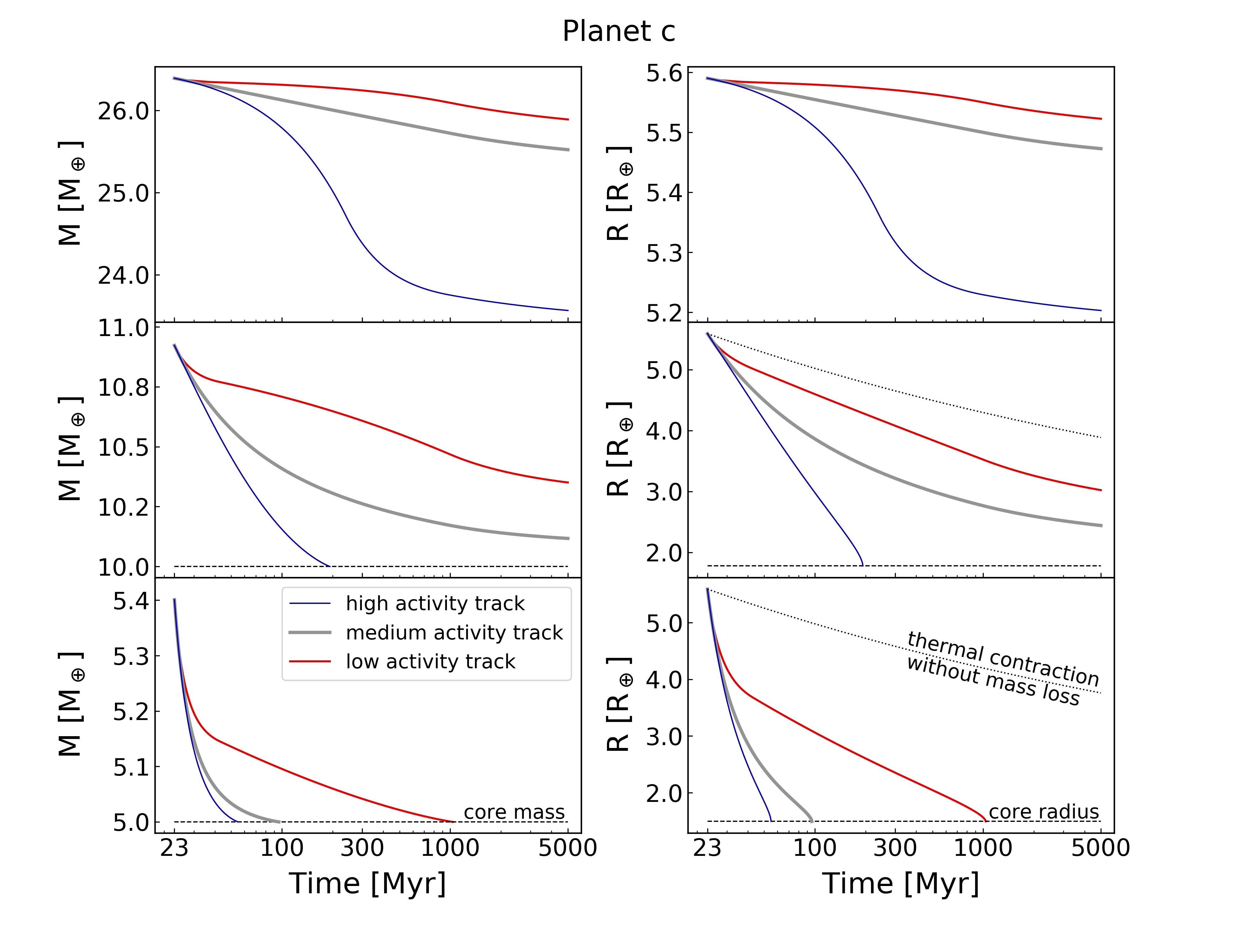}
\caption{Results of our calculations showing the possible future mass and radius evolution of the inner most planet c. The three panels on the left show the mass evolution for our ''high-density case'' planet (top) and the two ''fluffy planet cases'' with 10 and 5 Earth-mass cores (middle and bottom, respectively). The panels on the right show the corresponding radius evolutions. The red, grey and blue lines represent the planetary evolution considering a high, medium or low stellar activity evolution for V1298 Tau.}
\label{fig:evo_planet_c}
\end{figure}

\begin{figure}
\includegraphics[trim=30 25 30 0,clip,width=0.50\textwidth]{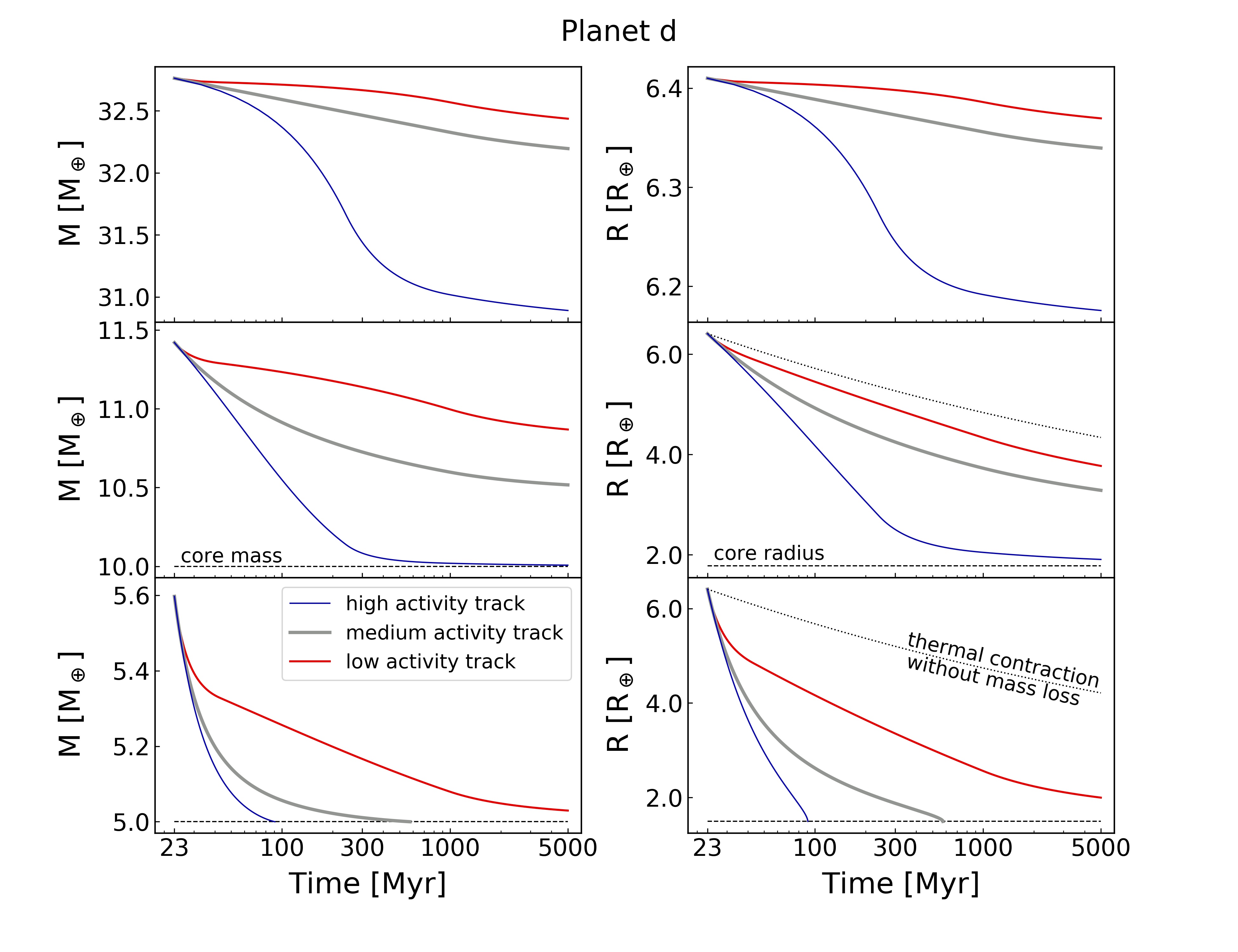}
\caption{Same as Fig. \ref{fig:evo_planet_c}, but for planet d.}
\label{fig:evo_planet_d}
\end{figure}

We perform these calculations for all three stellar activity evolution tracks (see Fig. \ref{fig:evo_tracks}). We tested different time step sizes and found that a step size of 1 Myr yields practically the same results for the ''high density scenario'' planets as for smaller time steps. Due to the much faster radius evolution for planets in the ''fluffy-planet scenario'', we chose a smaller step size of 0.1 Myr. The planetary mass and radius evolution for the two innermost planets c and d are shown in Figures \ref{fig:evo_planet_c} and \ref{fig:evo_planet_d}.


\begin{figure*}
\includegraphics[width=0.95\textwidth]{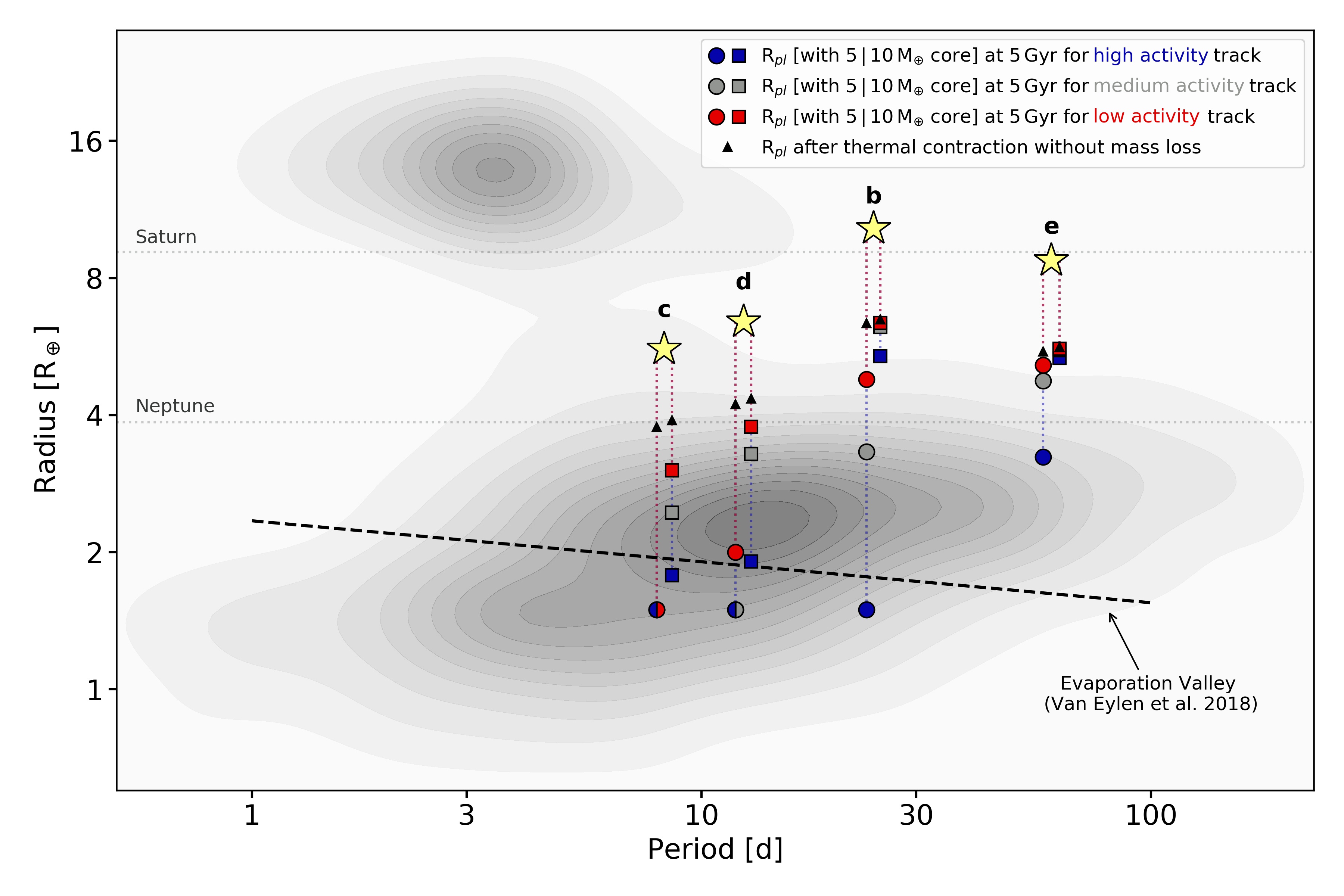}
\caption{Current position of the four transiting planets around V1298 Tau in the radius-period diagram. The gold stars mark the radii and periods measured by David et al. (2019) at a system age of $\sim23\,\mathrm{Myr}$. The vertical dotted lines show possible evolutionary tracks under atmospheric photoevaporation. The dots and squares mark the planetary radii at an age of 5 Gyr resulting from our calculations for planets with a $5\,$M$_\oplus$ and  $10\,$M$_\oplus$ core, respectively. The colours correspond to the stellar activity tracks assumed in the calculation, with blue colours corresponding to the high activity track, grey to the medium activity track and red to the low activity track.}
\label{fig:radiusgap}
\end{figure*}

Our calculations show that the high-energy evolution of the star has a significant effect on the planetary mass loss. The effect is particularly strong for ''fluffy'' scenario of close-in planets c and d, as they have extended atmospheres and are exposed to much higher XUV fluxes, see Fig.~\ref{fig:evo_planet_c} and \ref{fig:evo_planet_d}. In the stellar high activity scenario, those two planets can lose all of their modelled gaseous envelope, which corresponds to a lost mass fraction of around 10\%.

If we assume that the host star follows the evolutionary track for a slowly rotating star (low activity track), and will spend no more time in the saturation regime, the photoevaporative mass loss is, especially for the two outer planets, less severe. 
A summary of the possible radius changes and the remaining envelope mass fractions is given in Table~\ref{tab:mass_evo}.

For the planets in the ''high density scenario'' our calculations show that such high-mass planets would undergo little mass and radius evolution even for high activity track, owing to the larger planetary gravitational potential and the resulting lower mass loss rates. Planets in the low activity scenario undergo negligible mass and radius evolution. 

To put these numbers in context with the observed radius gap in the exoplanet population, we show the initial planetary radii and those at a system age of 5 Gyr in Fig.~\ref{fig:radiusgap}. The figure only show the evolution of the two ''fluffy planet'' scenarios with different core masses, as the ''high density'' scenario shows very little radius evolution. The combined impact of the stellar activity evolution and the planetary core mass can be seen particularly well for planets d and b, as their evolution on the period-radius diagram shows them either crossing or not crossing the photoevaporation valley around 2 R$_\oplus$ depending on whether V1298 Tau will follow the high, medium or low activity track.

\subsection{Uncertainties and model limitations}\label{limitations}

Some uncertainties are introduced into our modelling from observed or assumed parameters, others from the underlying assumptions of the model. We discuss the influence of these uncertainties here briefly.

The measured X-ray flux has an uncertainty of the order of 10\%, which makes very little difference in our modelling of the planet evolution. A 10\% higher or lower starting point of the stellar high-energy flux translates to a difference in the final planet radii of the order of 1\% in cases where the planet envelopes do not fully evaporate, and a difference in age where full envelope evaporation occurs of the order of 10\% for the other cases.

We have assumed an evaporation efficiency parameter $\epsilon$ of 0.1, which is not unusual in comparison to the existing literature, where a wide range of values has been employed in the past. However, this also needs to be seen in the context of applicability of the energy-limited escape scenario itself. The regime of energy-limited escape has been modelled to be valid for low to moderate XUV fluxes onto the planetary atmosphere, and is consistent with observations in the solar system \citep{Watson1981}.
For high XUV fluxes, it is expected that the energy supplied by the XUV photons does no longer go mainly into lifting the planetary atmosphere out of the planet's gravitational well. Instead, the planetary atmosphere is heated to such high temperatures that it starts to cool through emission lines, which reduces the energy available for expanding the atmosphere upwards. The escape rates could then be suppressed by a factor of up to ten compared to an $\epsilon$ factor appropriate for an energy-limited situation \citep{Murray-Clay2009}. We currently do not take into account such deviations from the energy-limited scenario, but assume on overall time-constant efficiency parameter of $\epsilon=0.1$ instead.

Also currently not included in our model are hydrodynamic or magnetic effects, such as a stellar wind streaming around the planet or the potential shielding effect that a planetary magnetosphere could provide against evaporation. Both are expected to influence the total mass loss. The stellar wind may enhance the mass loss of exoplanets, especially if the planets have no (or only a weak) magnetic field \citep{Cohen2015, Dong2017}. However, the existence of a planetary magnetosphere may lower the evaporative mass loss, particularly from the night side of the planet \citep{Owen2014}. Planetary magnetic fields in the solar system display a relationship between the angular momentum of the planetary spin and their observed magnetic field strength (see \citet{Griessmeier2004} and references therein), meaning that heavy, quickly spinning planets have strong magnetic fields. If the V1298~Tau planets are heavy and not fluffy, and are spinning fast -- which may be reasonable to assume since they are young --, magnetic shielding could be relevant to the mass loss.

\normalfont
In summary, our evaporation model is relatively simple and does not attempt to include all potentially relevant physical aspects of exoplanet evaporation, such as stellar winds, magnetic shielding, or any hydrodynamic effects. 
However, it is still instructive to see how even the inclusion of relatively few physical parameters, especially the stellar activity evolution and the planetary mass, can already cause a wide variety of possible future evolution tracks for the planets. In this context it is particularly important to measure masses of exoplanets around young stars, even though this is challenging, so that at least the planetary mass parameter in evaporation models can be constrained usefully.

\begin{table*}
\centering
\caption{Estimates for the present-day mass loss rates and planetary radii, as well as radii at 5 Gyr for two ''fluffy'' and one ''high-density'' planet scenarios given the three stellar activity evolutionary tracks. For the ''fluffy'' cases, core masses need to be explicitly assumed and we also report the planets' envelope mass fractions.}
\label{tab:mass_evo}
\begin{tabular}{lrrrrr}
\hline \hline 

& present-day  & present-day  &  \multicolumn{3}{c}{$R_p$ [R$_\oplus$] (envelope mass fraction $f_\mathrm{env}$ [\%]) at 5 Gyr  } \\ [0.1cm] \cline{4-6} \noalign{\smallskip} 
Scenario & $\dot{M}$ [\gs] &  $R_p$ [R$_\oplus$] ($f_\mathrm{env}$ [\%]) & high activity track & medium activity track & low activity track\\ \hline
Planet c [M$_{\mathrm{core}}=5$\,$M_\oplus$]   &   $1.5\times10^{13}$     & 5.6 (7.4)   & 1.5 (0.0)     & 1.5 (0.0)      & 1.5 (0.0)\\
Planet c [M$_{\mathrm{core}}=10$\,$M_\oplus$]  &   $5.3\times10^{12}$     & 5.6 (8.5)   & 1.8 (0.0)     & 2.4 (1.2)      & 3.0 (3.5)\\
Planet c [high-density]                        &   $1.5\times10^{12}$     & 5.6 (--)    & 5.2 (--)      & 5.5 (--)           & 5.5 (--)\\ [0.1cm] 
Planet d [M$_{\mathrm{core}}=5$\,$M_\oplus$]   &   $1.2\times10^{13}$     & 6.4 (10.7)   & 1.5 (0.0)     & 1.5 (0.0)      & 2.0 (0.6)\\
Planet d [M$_{\mathrm{core}}=10$\,$M_\oplus$]  &   $4.4\times10^{12}$     & 6.4 (12.4)   & 1.9 (0.1)     & 3.3 (4.9)      & 3.8 (8.0)\\
Planet d [high-density]                        &   $9.7\times10^{11}$     & 6.4 (--)     & 6.2  (--)     & 6.3 (--)       & 6.4 (--) \\ [0.1cm] 
Planet b [M$_{\mathrm{core}}=5$\,$M_\oplus$]   &   $1.6\times10^{13}$     & 10.3 (33.9)  & 1.5 (0.0)     & 3.3 (5.7)      & 4.8 (16.2)\\
Planet b [M$_{\mathrm{core}}=10$\,$M_\oplus$]  &   $4.6\times10^{12}$     & 10.3 (43.3)  & 5.4 (25.2)    & 6.2 (38.6)     & 6.4 (41.2)\\
Planet b [high-density]                        &   $6.6\times10^{11}$     & 10.3 (--)    & 10.2 (--)     & 10.2 (--)      & 10.2 (--)\\ [0.1cm] 
Planet e [M$_{\mathrm{core}}=5$\,$M_\oplus$]   &   $2.7\times10^{12}$     & 8.7 (25.7)  & 3.2 (5.8)     & 4.8 (17.4)     & 5.2 (21.6)\\
Planet e [M$_{\mathrm{core}}=10$\,$M_\oplus$]  &   $9.0\times10^{11}$     & 8.7 (31.7)  & 5.3 (27.1)    & 5.6 (30.4)     & 5.6 (31.0)\\
Planet e [high-density]                        &   $1.5\times10^{11}$     & 8.7 (--)    & 8.7 (--)      & 8.7 (--)       & 8.7 (--)\\
\hline
\end{tabular}
\end{table*}

\section{Conclusions}

We use X-ray observations of the young exoplanet host star V1298 Tau with Chandra and ROSAT to estimate the current high-energy irradiation the four Neptune- to Saturn-sized planets are exposed to. We find that V1298 Tau, with an age of $\sim$ 23 Myr, is X-ray bright with a luminosity of $\log L_X$\,[\lumi] $=30.1$, and has a mean coronal temperature of approximately 9 MK. By employing a model for the stellar activity evolution together with exoplanetary mass loss we then estimate the atmospheric evolution of the four planets. Due to the lack of measured masses, it is challenging to provide constraining predictions on the fate of these four very young planets. We therefore estimate the planetary mass and radius evolution for a ''fluffy-planet scenario'' and a ''high-density scenario'', covering a realistic/conceivable mass range. We model the four planets as fluffy planets with a 5 and 10 M$_\oplus$ rocky core underneath a thick hydrogen/helium envelope, and also as four higher-mass/density planets with masses ranging roughly between those of Neptune and Saturn ($\sim$ 20-70 M$_\oplus$). We show that, as expected, the low-mass planets are most affected by photoevaporative mass loss, mainly due to their weaker gravitational potential and the consequently higher mass-loss rates. Our results show that the stellar activity evolution and the age at which spin-down sets in can make a significant difference in possible life-time evaporation outcomes for the planets.

\section*{Acknowledgements}

The scientific results reported in this article are based in part on observations made by the Chandra X-ray Observatory, and in part based on data from the ROSAT Data Archive of the Max-Planck-Institut f\"ur extraterrestrische Physik (MPE) at Garching, Germany. This research made use of Astropy,\footnote{http://www.astropy.org} a community-developed core Python package for Astronomy \citep{astropy:2013, astropy:2018}. 
Parts of this work was supported by the German \emph{Leibniz-Gemeinschaft}, project number P67-2018.




\bibliographystyle{mnras}
\bibliography{paper_bib} 



\appendix

\section{Planet evolution plots}

Large versions of the radius and mass evolution plots for all four planets are displayed in this appendix.

\begin{figure*}
\includegraphics[width=0.8\textwidth]{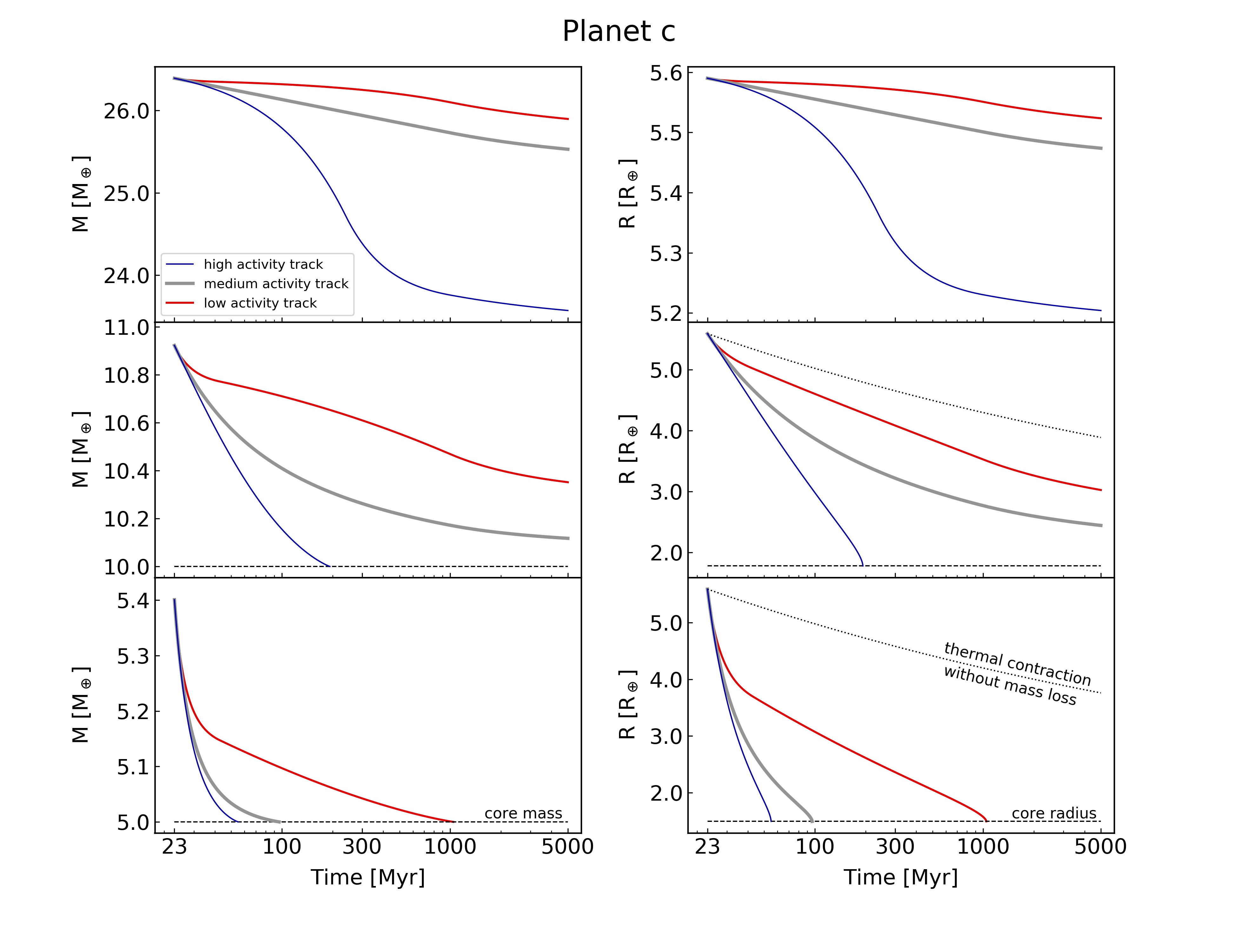}
\caption{High-resolution version of Fig.~\ref{fig:evo_planet_c}.}
\label{fig:app_evo_planet_c}
\end{figure*}

\begin{figure*}
\includegraphics[width=0.8\textwidth]{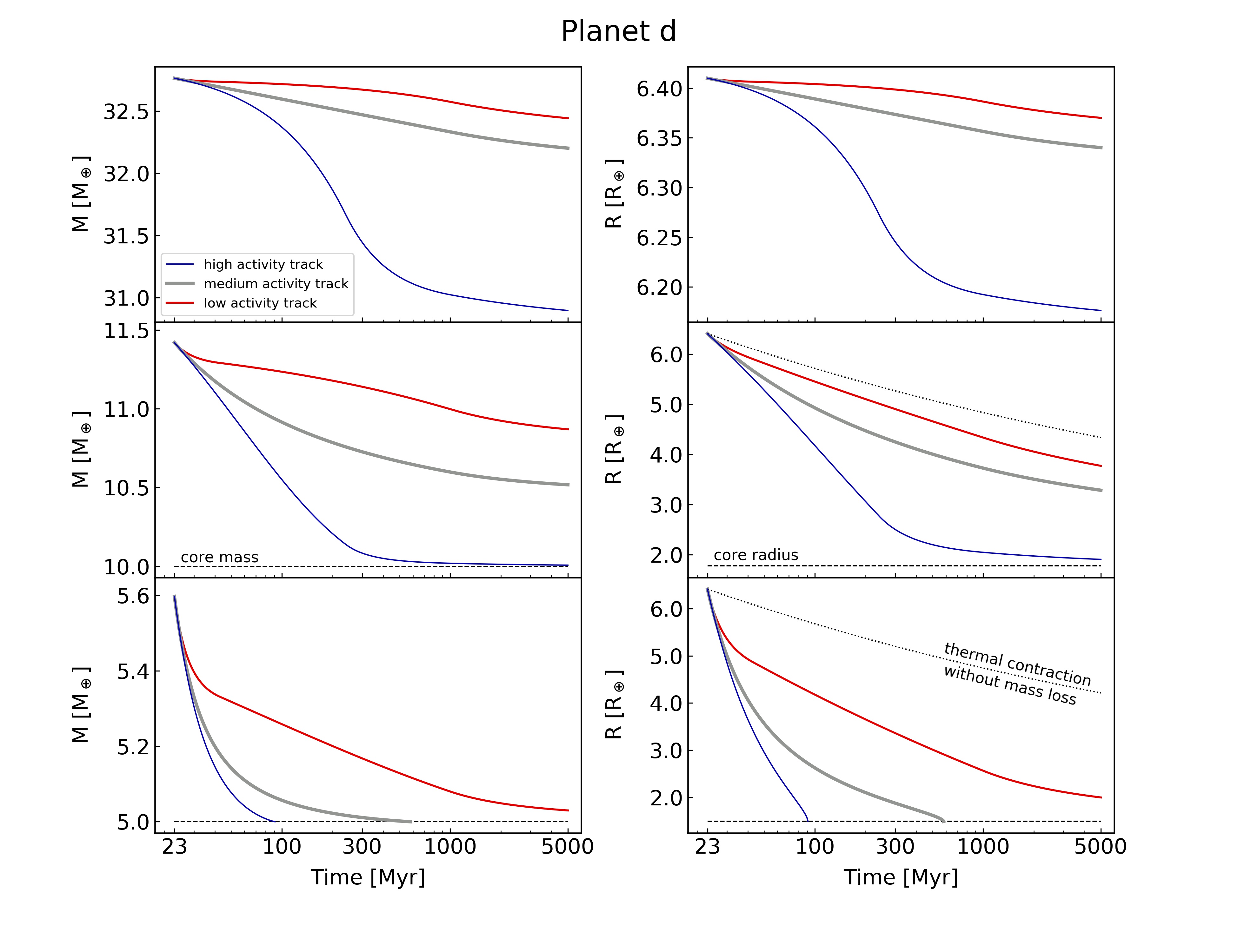}
\caption{High-resolution version of Fig.~\ref{fig:evo_planet_d}.}
\label{fig:app_evo_planet_d}
\end{figure*}

\begin{figure*}
\includegraphics[width=0.8\textwidth]{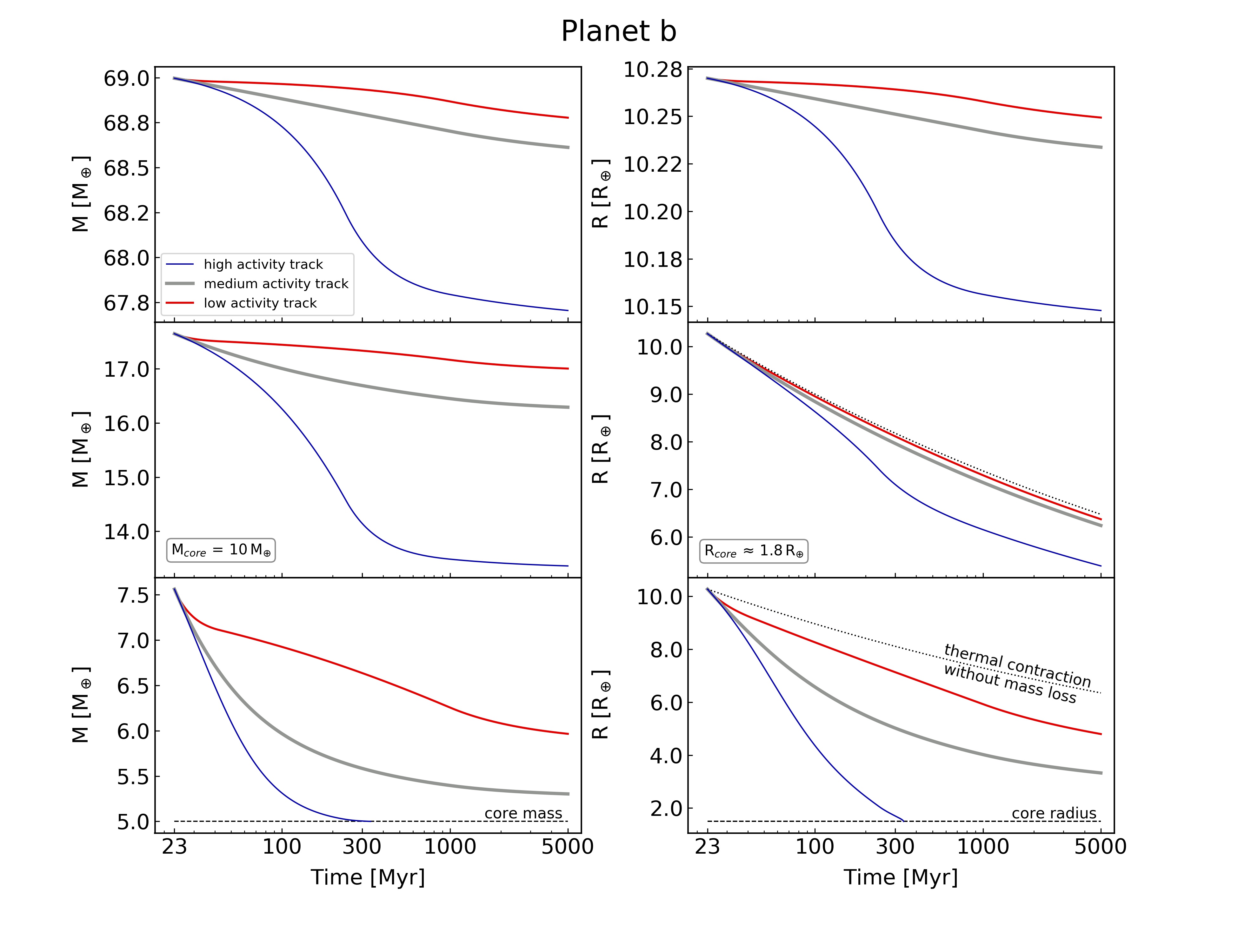}
\caption{Same as Fig.~\ref{fig:evo_planet_c}, but for planet b.}
\label{fig:app_evo_planet_b}
\end{figure*}

\begin{figure*}
\includegraphics[width=0.8\textwidth]{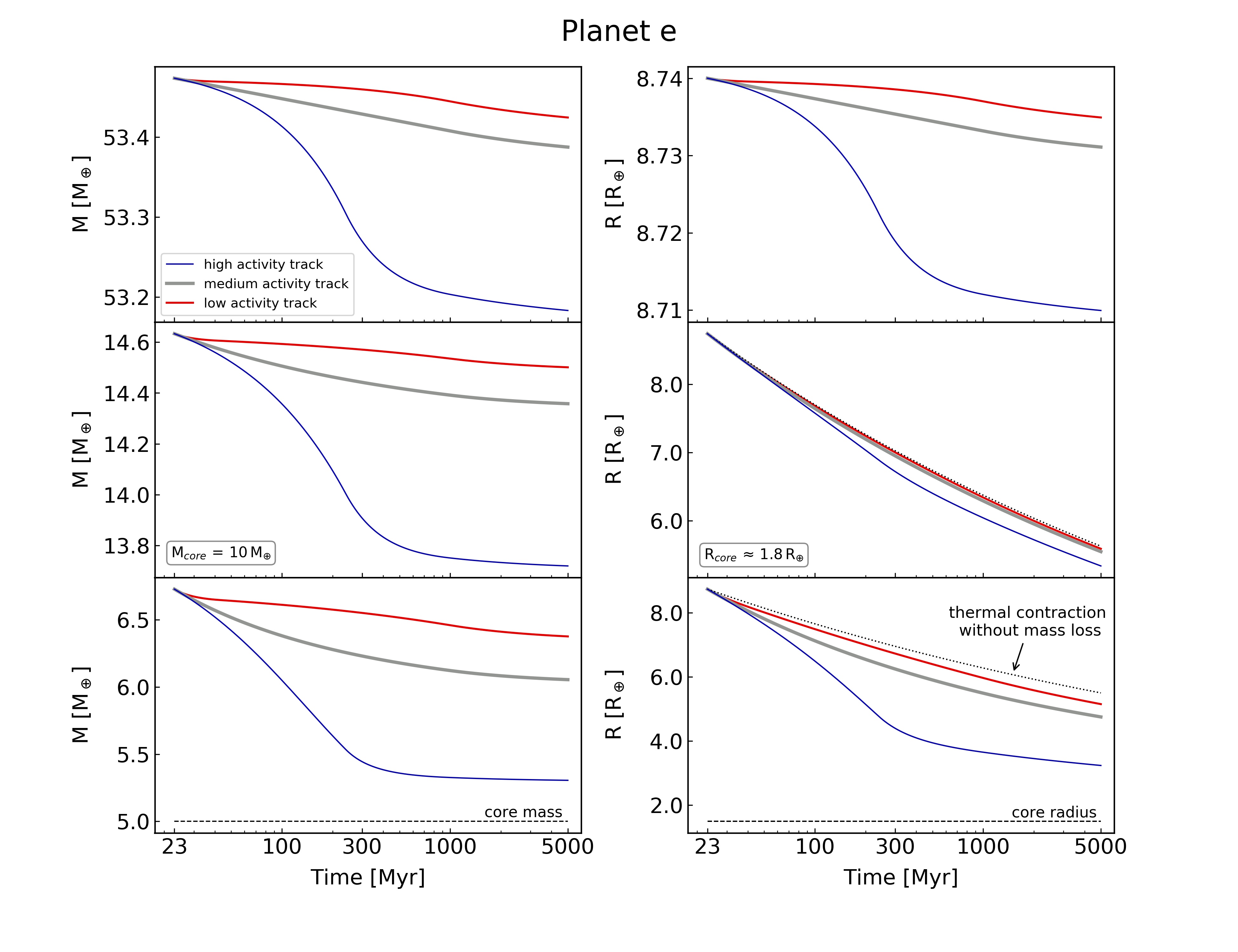}
\caption{Same as Fig.~\ref{fig:evo_planet_c}, but for planet e.}
\label{fig:app_evo_planet_e}
\end{figure*}




\bsp	
\label{lastpage}
\end{document}